\documentclass[aps,prc,showpacs,twocolumn,groupedaddress]{revtex4}

\usepackage{graphics}

\begin{document}

%\preprint{}

\title{Structure of cold nuclear matter at subnuclear densities
  by Quantum Molecular Dynamics}


\author{Gentaro Watanabe$^{a,b}$, Katsuhiko Sato$^{a,c}$,
  Kenji Yasuoka$^{d}$ and Toshikazu Ebisuzaki$^{b}$}
%\thanks{}
\affiliation{
$^{a}$Department of Physics, University of Tokyo,
Tokyo 113-0033, Japan
\\
$^{b}$Division of Computational Science, RIKEN,
Saitama 351-0198, Japan
\\
$^{c}$Research Center for the Early Universe, 
University of Tokyo,
Tokyo 113-0033, Japan
\\
$^{d}$Department of Mechanical Engineering, Keio University,
Yokohama 223-8522, Japan}


\date{\today}

\begin{abstract}
Structure of cold nuclear matter
at subnuclear densities for the proton fraction $x=0.5$, 0.3 and 0.1
is investigated by quantum molecular dynamics (QMD) simulations.
We demonstrate that the phases with slablike and
rodlike nuclei, etc. can be formed dynamically from hot uniform nuclear matter
without any assumptions on nuclear shape, and also systematically
analyze the structure of cold matter using two-point correlation functions
and Minkowski functionals.
In our simulations, we also observe intermediate phases,
which has complicated nuclear shapes.
It is found out that these phases
can be characterized as those with negative Euler characteristic.
Our result implies the existence of these kinds of phases
in addition to the simple ``pasta'' phases
in neutron star crusts and supernova inner cores.

In addition, we investigate the properties of the effective QMD interaction
used in the present work to examine the validity of our results.
The resultant energy per nucleon $\epsilon_{n}$ of the pure neutron matter,
the proton chemical $\mu_{p}^{(0)}$ in pure neutron matter and
the nuclear surface tension $E_{\rm surf}$ are generally reasonable
in comparison with other nuclear interactions.

\end{abstract}

%\keywords{}

\maketitle

\section{Introduction}

For the past several decades since the discovery of pulsars,
many authors have investigated the properties of dense matter
which exist inside neutron stars and supernova cores
(see, e.g., Refs.\ \cite{bbs,bbp,sato}).
These objects have been shown that
they consist of a variety of material phases
whose physical properties reflect in many astrophysical phenomena
of these objects.
Especially, the properties of nuclear matter under extreme conditions,
which is one of the essential topics for understanding the mechanism of
collapse-driven supernovae \cite{bethe},
the structure of neutron star crusts \cite{review} and its relating phenomena,
have been studied actively.
This subject is also interesting
as one of the fundamental problems of the complex fluids of nucleons.

At subnuclear densities,
nuclear matter exhibits the coexistence of a liquid phase with a gas phase
due to the internucleon interaction which has an attractive part.
At sufficiently low temperatures
relevant to neutron star interiors,
and sufficiently below the normal nuclear density,
long-range Coulomb interactions make the system divide periodically
into gas and spherical liquid drops,
adding a crystalline property to the liquid-gas coexistence.

In the density region
where nuclei are about to melt into uniform matter,
it is expected that the energetically favorable configuration
of the mixed phase possesses interesting spatial structures
such as rodlike and slablike nuclei and rodlike and spherical bubbles, etc.,
which are referred to as nuclear ``pasta''.
This picture was originally proposed
by Ravenhall et al. \cite{rpw}
and Hashimoto et al. \cite{hashimoto} independently.
Their predictions both based on free energy calculations
with liquid drop models assuming some specific nuclear shapes.
These works clarify that
the most energetically stable nuclear shape is
determined by a subtle balance
between the nuclear surface and Coulomb energies.
Detailed aspects of equilibrium phase diagrams,
such as a series of nuclear shapes
which can be realized as the energetically most favorable and
the density range corresponding to the phases with nonspherical nuclei,
vary with nuclear models \cite{lorenz}.
However, the realization of the ``pasta'' phases as energy minimum states
can be seen in a wide range of nuclear models and
the phase diagrams possess an universal basic feature that,
with increasing density,
the shape of the nuclear matter region changes like
\ sphere $\rightarrow$ cylinder $\rightarrow$ slab $\rightarrow$
cylindrical hole $\rightarrow$ spherical hole $\rightarrow$ uniform
\cite{gentaro1,gentaro2}.
%%%
%basic feature of phase diagrams is universal \cite{gentaro1,gentaro2},
%i.e. with increasing density,
%the shape of the nuclear matter region changes like
%\ sphere $\rightarrow$ cylinder $\rightarrow$ slab $\rightarrow$
%cylindrical hole $\rightarrow$ spherical hole $\rightarrow$ uniform.
%%%
This feature is also reproduced by the Thomas-Fermi calculations
by several groups \cite{williams,lassaut,oyamatsu}.

The phases with these exotic nuclear structures,
if they were realized in neutron star crusts or supernova cores,
bring about many astrophysical consequences.
As for those in neutron star phenomena,
it is interesting to note the relevance of nonspherical nuclei
to pulsar glitches and cooling of neutron stars.
Although the question whether the mechanism of pulsar glitches
is depicted by vortex pinning model or star quake model
has yet to be settled completely,
the existence of nonspherical nuclei in neutron star matter (NSM)
have significant effects in both cases.
As for the former, while the force needed to pin vortices
has yet to be clarified completely even for a bcc lattice of spherical nuclei
mainly due to the uncertain properties of impurities and defects \cite{jones},
the effect of spatial structure of normal nuclear matter
on vortex dynamics cannot be ignored.
As for the latter, the existence of ``pasta'' phases
with slablike and rodlike nuclei would change the elastic properties of
inner crust matter from those of crystalline solid to those of liquid crystal
as indicated by Pethick and Potekhin \cite{pp},
which results in significant decrease of the maximum elastic energy
that can be stored in the inner crust.
The presence of nonspherical nuclei would also accelerate
the cooling of the corresponding region of neutron stars
by opening semileptonic weak processes which are unlikely to occur
for spherical nuclei \cite{lorenz}.

``Pasta'' phases in supernova matter (SNM) are expected to affect
the neutrino transport and hydrodynamics in supernova cores.
Let us first note that
the neutrino wavelengths, typically of order 20 fm, 
are comparable to or even greater than the internuclear spacing,
leading to diffractive effects on the neutrino elastic scattering
off such a periodic spatial structure of nuclear matter \cite{rpw}.
These effects, induced by the internuclear Coulombic correlations,
would reduce the scattering rates and hence the lepton 
fraction $Y_{L}$.
For the bcc lattice of spherical nuclei, such a reduction was
examined by Horowitz \cite{horowitz} by calculating the 
associated static structure factor.
It is also noteworthy that 
nonspherical nuclei and bubbles are elongated in specific direction.
In such direction, the neutrino scattering processes are
no longer coherent,
in contrast to the case of
roughly spherical nuclei whose finiteness in any direction
yields constructive interference in the scattering.
The final point to be mentioned is that
the changes in the nuclear shape are accompanied by
discontinuities in the adiabatic index,
denoting how hard the
equation of state of the material is.
These discontinuities
may influence the core hydrodynamics during the initial phase of 
the collapse \cite{lassaut}.

Though the properties of ``pasta'' phases in equilibrium state
have been investigated actively,
the formation and the melting processes of these phases have not been discussed
except for some limited cases
which are based on perturbative approaches \cite{review,iida}.
It is important to adopt a microscopic and dynamical approach
which allows arbitrary nuclear structures
in order to understand these processes of nonspherical nuclei.
At finite temperatures, it is considered that
not only nuclear surface becomes obscure but also nuclei of various shapes
may coexist.
Therefore, it is necessary to incorporate density fluctuations
without any assumptions on nuclear shape
to investigate the properties of ``pasta'' phases at finite temperatures.
Although the works done by Williams and Koonin \cite{williams}
and Lassaut et al. \cite{lassaut} do not assume nuclear structure,
they can not incorporate fluctuations of nucleon distributions
in a satisfying level
because they are based on the Thomas-Fermi calculation,
which is one-body approximation.
In addition, only a single structure is contained in the simulation box
in these works,
there are thus possibilities that nuclear shape is strongly affected
by boundary effect and some structures are prohibited implicitly.

In the present work, we study the structure of cold dense matter
at subnuclear densities in the framework of
quantum molecular dynamics (QMD) \cite{aichelin},
which is one of the molecular dynamics (MD) approaches
for nucleon many-body systems (see, e.g., Ref. \cite{feldmeier} for review).
MD for nucleons including QMD, which is a microscopic and dynamical method
without any assumptions on nuclear structure,
is suitable for incorporating fluctuations of particle distributions.
Previously, we have reported the first results of our study
on nuclear ``pasta'' by QMD, which demonstrated that the ``pasta'' phases
can be formed in a dynamical way for matter with proton fractions
$x=0.3$ and 0.5 \cite{qmd1}.
In this full paper, we present new results for
astrophysically interesting neutron-rich matter of $x=0.1$
in addition to the cases of $x=0.3$ and 0.5 reported before.

The plan of this paper is as follows.
In Section II, we describe the framework of the QMD model
used in the present study.  We then show the results of our simulations
in Section III and analyze the structure of matter obtained by the simulations
using two-point correlation functions and Minkowski functionals in Section IV.
In Section V, we investigate the properties of
the effective nuclear interaction used in this work
in order to examine the validity of our results in terms of nuclear forces.
Astrophysical discussions are given in Section VI.
Summary and conclusions are presented in Section VII.

%%%%%%%%%%%%%%%%%%%%%%%%%%%%%%%%%%%%
\section{Quantum Molecular Dynamics}

We have various types of molecular dynamics methods for nucleons
including representative ones such as
fermionic molecular dynamics (FMD) \cite{fmd},
antisymmetrized molecular dynamics (AMD) \cite{amd} and QMD, etc.
In the present work, we choose QMD among them
balancing between calculation cost and accuracy.
The typical length scale $l$ of inter-structure
is $l \sim 10$ fm and the density region of interest is
just below the normal nuclear density $\rho_{0} = 0.165 {\rm\ fm}^{-3}$.
The required nucleon number $N$ in order to reproduce $n$ unit structures
in the simulation box is about $N \sim \rho_{0} (n l)^{3}$ (for slabs).
It is thus desirable that we prepare nucleons of order 10000
if we try to reduce boundary effects down to a satisfactory level
by reproducing several several unit structures in the box.
While it is a hard task to treat such a large system
with, for example, FMD and AMD
whose calculation costs scale as $\sim N^{4}$,
it is feasible to do it with QMD whose calculation costs scale
as $\sim N^{2}$.  This difference comes from summations
in the Slater determinants in the trial wave functions of the former models.
In QMD, on the other hand, the total $N$-nucleon wave function
$|\Phi\rangle$ is assumed to be a direct product of single-nucleon states
$|\phi_{i}\rangle$ :
\begin{equation}
  |\Phi\rangle = |\phi_{1}\rangle \otimes |\phi_{2}\rangle \otimes
  \cdots \otimes |\phi_{N}\rangle\ .\label{trial func}
\end{equation}
The single-nucleon state is represented by a Gaussian wave packet:
\begin{equation}
  \phi_{i}({\bf r}) = \langle{\bf r}|\phi_{i}\rangle
     = \frac{1}{(2\pi L)^{3/4}}
     \exp{\left[ -\frac{({\bf r}-{\bf R}_{i})^{2}}{4 L}
       +\frac{i}{\hbar}\ {\bf r}\cdot{\bf P}_{i} \right]}\ ,\label{packet}
\end{equation}
where ${\bf R}_{i}(t)$ and ${\bf P}_{i}(t)$ are the centers of position
and momentum of the packet $i$, respectively,
and $L$ is a parameter related to the extension of the wave packet
in the coordinate space.

It is also noted that we mainly focus on the macroscopic structures;
the exchange effect would not be so important for them.
This can be seen by comparing the typical values of
the exchange energy for the macroscopic scale
and of the energy difference
between two successive phases with nonspherical nuclei.
Suppose there are two identical nucleons, $i=1$ and 2, bound in
different nuclei each other.
The exchange energy between these particles is calculated
as an exchange integral:
\begin{equation}
K= \int U({\bf r}_{1}-{\bf r}_{2})\
\varphi_{1}({\bf r}_{1}) \varphi_{1}^{*}({\bf r}_{2})
\varphi_{2}({\bf r}_{2}) \varphi_{2}^{*}({\bf r}_{1})\
d{\bf r}_{1} d{\bf r}_{2}\ ,
\end{equation}
where $U$ is the potential energy.
An asymptotic form of the wave function is given by
\begin{equation}
\varphi_{i} \sim \exp{(- \kappa_{i} r)}\ ,
\end{equation}
with $r = |{\bf r}_{1}-{\bf r}_{2}|$ and
$\kappa_{i} = \frac{1}{\hbar} \sqrt{2mE_{i}}\ ,\ ( i=1,2 )$,
where $E_{i}$ is the binding energy and $m$ is the nucleon mass.
The exchange integral reads
\begin{equation}
K \sim \exp{[-(\kappa_{1}+\kappa_{2})R]} \sim 5 \times 10^{-6} {\rm MeV}
\end{equation}
for the internuclear distance $R \simeq 10$ fm
and $E_{i} \simeq 8$ MeV,
which is extremely smaller than the typical energy difference
per nucleon between the ``pasta'' phases
of order 0.1 keV (for NSM, see Fig.\ 4 in Ref.\ \cite{gentaro1}) - 10 keV
(for SNM, see Fig.\ 4 in Ref.\ \cite{gentaro2}).
Therefore, it is expected that QMD, which is less elaborate
in treating the exchange effect, is not bad approximation
for investigating the nuclear ``pasta''.
Consequently, QMD has the advantages over the other models
in the present study.
In the future, we will have to confirm the validity of the results
obtained by QMD using other more elaborate model such as AMD or FMD
to treat the exchange effect more precicely.
%%%%%
%by some other model which is more elaborate
%in treating the exchange effect such as AMD or FMD,
%but this problem is beyond the scope of the present work.
%%%%%
However, this problem is beyond the scope of the present work.

\subsection{Model Hamiltonian}

To simulate nuclear matter at subnuclear densities within the framework of QMD,
we use a QMD model Hamiltonian developed by Maruyama et al. \cite{maruyama},
which is constructed so as to reproduce bulk properties of nuclear matter
and properties of finite nuclei.
This model Hamiltonian consists of the following six terms
\begin{equation}
  {\cal H} = 
  T+V_{\rm Pauli}+V_{\rm Skyrme}+V_{\rm sym}+V_{\rm MD}+V_{\rm Coulomb}\ ,
  \label{hamiltonian}
\end{equation}
where $T$ is the kinetic energy,
$V_{\rm Pauli}$ is the Pauli potential introduced to reproduce
the Pauli principle effectively,
$V_{\rm Skyrme}$ is the Skyrme potential
which consists of an attractive two-body term and a repulsive three-body term,
$V_{\rm sym}$ is the symmetry potential,
$V_{\rm MD}$ is the momentum-dependent potential
introduced as two Fock terms of the Yukawa interaction and
$V_{\rm Coulomb}$ is the Coulomb potential.
The expressions of these terms are given as
\begin{widetext}
\begin{eqnarray}
  T  &=& \sum_{i, j(\ne i)} \frac{\bf P_{\it i}^{2}}{2 m_{i}}\ ,\label{kin}\\  
  V_{\rm Pauli} &=& 
  \frac{1}{2}\
  C_{\rm P}\left( \frac{\hbar}{q_0 p_0}\right)^3
  \sum_{i, j(\neq i)} 
  \exp{ \left [ -\frac{({\bf R}_i-{\bf R}_j)^2}{2q_0^2} 
          -\frac{({\bf P}_i-{\bf P}_j)^2}{2p_0^2} \right ] }\
  \delta_{\tau_i \tau_j} \delta_{\sigma_i \sigma_j}\ ,\label{pauli}\\
  V_{\rm Skyrme} &=&
  {\alpha\over 2\rho_0}\sum_{i, j (\neq i)}
  \rho_{ij}
  +  {\beta\over (1+\tau)\ \rho_0^{\tau}}
  \sum_i \left[ \sum_{j (\neq i)} \int { d^3{\bf r} \ \tilde{\rho_i}({\bf r}) \
                       \tilde{\rho_j}({\bf r}) } \right]^{\tau}\ ,
                   \label{skyrme}\\
  V_{\rm sym} &=&
  {C_{\rm s}\over 2\rho_0} \sum_{i , j(\neq i)} \,
  ( 1 - 2 | c_i - c_j | ) \ \rho_{ij}\ ,\label{sym}\\
  V_{\rm MD}  &=&
  V_{\rm MD}^{(1)} + V_{\rm MD}^{(2)} \nonumber \\
  &=&
         {C_{\rm ex}^{(1)} \over 2\rho_0} \sum_{i , j(\neq i)} 
      {1 \over 1+\left[{{\bf P}_i-{\bf P}_j \over \hbar \mu_1}\right]^2} 
      \ \rho_{ij}
     +   {C_{\rm ex}^{(2)} \over 2\rho_0} \sum_{i , j(\neq i)} 
      {1 \over 1+\left[{{\bf P}_i-{\bf P}_j \over \hbar \mu_2}\right]^2} 
      \ \rho_{ij}\ ,\label{md}\\
  V_{\rm Coulomb} &=&
  {e^2 \over 2}\sum_{i , j(\neq i)}
  \left(\tau_{i}+\frac{1}{2}\right) \, \left(\tau_{j}+\frac{1}{2}\right)
  \int\!\!\!\!\int d^3{\bf r}\,d^3{\bf r}^{\prime} 
  { 1 \over|{\bf r}-{\bf r}^{\prime}|} \,
  \rho_i({\bf r})\rho_j({\bf r}^{\prime})\ ,\label{coulomb}
\end{eqnarray}
\end{widetext}
where $\rho_{ij}$ means the overlap between the single-nucleon densities,
$\rho_{i}({\bf r})$ and $\rho_{j}({\bf r})$,
for $i$-th and $j$-th nucleons given as
\begin{equation}
  \rho_{ij} \equiv \int { d^3{\bf r} \ \rho_i({\bf r}) \
                       \rho_j({\bf r}) }\ ,
\end{equation}
$\sigma_{i}$ is the nucleon spin and $\tau_{i}$ is the isospin
($\tau_{i}=1/2$ for protons and $-1/2$ for neutrons) and
$C_{\rm P},\ q_{0},\ p_{0},\ \alpha,\ \beta,\ \tau,\ C_{\rm s},\ C_{\rm ex}^{(1)},\ C_{\rm ex}^{(2)},\ \mu_{1},\ \mu_{2}$
and $L$ are model parameters
determined to reproduce the properties of the ground states
of the finite nuclei, especially heavier ones,
and the saturation properties of nuclear matter \cite{maruyama}.
A parameter set used in this work is shown in Table \ref{parameter}.
The single-nucleon densities $\rho_{i}({\bf r})$
and $\tilde{\rho_{i}}({\bf r})$
are given by
\begin{eqnarray}
  \rho_i({\bf r}) & = & \left| \phi_{i}({\bf r}) \right|^{2}
  = \frac{1}{(2\pi L)^{3/2}}\ \exp{\left[
                - \frac{({\bf r} - {\bf R}_i)^2}{2L} \right]}\ ,\quad \\
  \tilde{\rho_i}({\bf r}) & = &
  \frac{1}{(2\pi \tilde{L})^{3/2}}\ \exp{\left[
                - \frac{({\bf r} - {\bf R}_i)^2}{2\tilde{L}} \right]}\ ,
\end{eqnarray}
with
\begin{equation}
  \tilde{L} = \frac{(1+\tau)^{1/ \tau}}{2}\ L\ .
\end{equation}
The modified width $\tilde{L}$ in $\tilde{\rho_{i}}({\bf r})$
is introduced in the three-body term of Skyrme interaction
[Eq. (\ref{skyrme})] to incorporate the effect of
the repulsive density-dependent term.

\begin{table}[h]
\caption{Effective interaction parameter set\\
  \qquad (incompressibility $K$=280 MeV; medium EOS model in Ref.\ \cite{maruyama})}
\begin{ruledtabular}
\begin{tabular}{cccc}
& $C_{\rm P}$ (MeV) &\qquad\qquad 207 &\\
& $p_{0}$ (MeV/$c$) &\qquad\qquad 120 &\\
& $q_{0}$ (fm) &\qquad\qquad 1.644 &\\
& $\alpha$ (MeV) &\qquad\qquad $-92.86$ &\\
& $\beta$ (MeV) &\qquad\qquad 169.28 &\\
& $\tau$ &\qquad\qquad 1.33333 &\\
& $C_{\rm s}$ (MeV) &\qquad\qquad 25.0 &\\
& $C_{\rm ex}^{(1)}$ (MeV) &\qquad\qquad $-258.54$ &\\
& $C_{\rm ex}^{(2)}$ (MeV) &\qquad\qquad 375.6 &\\
& $\mu_1$ (fm$^{-1}$) &\qquad\qquad 2.35 &\\
& $\mu_2$ (fm$^{-1}$) &\qquad\qquad 0.4 &\\
& $L$ (fm$^2$) &\qquad\qquad 2.1 &\\
\end{tabular}
\end{ruledtabular}
\label{parameter}
\end{table}

We adopt QMD equations of motion with friction terms
to simulate the dynamical relaxation:
\begin{equation}
\begin{array}{ccr}
  \dot{\bf R}_{i} &=&
     {\displaystyle \frac{\partial {\cal H}}{\partial {\bf P}_{i}}
       - \xi_{R} \frac{\partial {\cal H}}{\partial {\bf R}_{i}}}\ ,\\ \\
  \dot{\bf P}_{i} &=&
     {\displaystyle -\frac{\partial {\cal H}}{\partial {\bf R}_{i}}
       - \xi_{P} \frac{\partial {\cal H}}{\partial {\bf P}_{i}}}\ ,\label{qmdeom fric}
\end{array}
\end{equation}
where the friction coefficients $\xi_{R}$ and $\xi_{P}$ are positive definite,
which determine the relaxation time scale.
The relaxation scheme given by Eqs.\ (\ref{qmdeom fric})
is referred to as the steepest descent method and
it leads to the continuous decrease in ${\cal H}$ as
\begin{eqnarray}
  \frac{d {\cal H}}{dt} &=&
  \dot{\bf R}_{i} \cdot\frac{\partial {\cal H}}{\partial {\bf R}_{i}}
  +\dot{\bf P}_{i}\cdot\frac{\partial {\cal H}}{\partial {\bf P}_{i}}
  \nonumber\\
  &=&-\ \xi_{R} \left(\frac{\partial {\cal H}}{\partial {\bf R}_{i}}\right)^{2}
  - \xi_{P} \left(\frac{\partial {\cal H}}{\partial {\bf P}_{i}}\right)^{2}
  \leq 0\ .
\end{eqnarray}
Even though it is recognized that this method is not efficient,
it is expected that the dynamics given by Eqs.\ (\ref{qmdeom fric}) with
$\xi_{R}, \xi_{P} \ll 1$ deviates slightly in a short period from
the physically grounded dynamics given by
QMD equations of motion without the friction terms
[equations without the second terms
in the right-hand sides of Eqs.\ (\ref{qmdeom fric})],
which we would like to respect.

%%%%%%%%%%%%%%%%%%%%%%%%%%%%%%%%%%%%%%%%
\section{QMD Simulations of Cold Matter at Subnuclear Densities}

\subsection{QMD Simulations for $x=0.5$ and 0.3\label{sim p-rich}}

We have performed QMD simulations of an infinite $(n,p,e)$ system
with fixed proton fractions $x = $ 0.5 and 0.3
for various nucleon densities $\rho$
[the density region is (0.05 - 1.0) $\rho_{0}$].
We set 2048 nucleons (1372 nucleons in some cases) contained in
a cubic box which is imposed by the periodic boundary condition.
Throughout this paper, the numbers of the protons (neutrons) with
up-spin and with down-spin are equal.
The relativistic degenerate electrons which ensure the charge neutrality
are regarded as a uniform background
and the Coulomb interaction is calculated by the Ewald method
taking account of the Gaussian charge distribution of each wave-packet
(see Appendix \ref{ewald sum}).
This method enables us to efficiently sum up
contributions of long-range interactions
in a system with periodic boundary conditions.
For nuclear interaction, we use the effective Hamiltonian
developed by Maruyama et al. (medium EOS model) \cite{maruyama}
whose expressions are given in the last Section.

We first prepare an uniform hot nucleon gas
at $k_{B}T \sim 20$ MeV as an initial condition
equilibrated for $\sim 500 - 2000$ fm/$c$ in advance.
In order to realize the ground state of matter,
we then cool it down slowly for $O(10^{3}-10^{4})$ fm/$c$,
keeping the nucleon density constant
with the frictional relaxation method
[Eqs.\ (\ref{qmdeom fric})], etc. \cite{note nose}
until the temperature gets $\sim 0.1$ MeV or less.
Note that no artificial fluctuations are given in the simulation.

The QMD equations of motion with the friction terms
given by Eqs.\ (\ref{qmdeom fric})
are solved using the fourth-order Gear predictor-corrector method
in conjunction with multiple time step algorithm \cite{allen}.
Integration time steps $\Delta t$ are set to be adaptive
in the range of $\Delta t < 0.1-0.2$ fm/$c$
depending on the degree of convergence.
At each step, the correcting operation is iterated until
the error of position $\Delta r$ and the relative error of momentum
$\Delta p/p$ become smaller than $10^{-6}$,
where $\Delta r$ and $\Delta p/p$ are estimated
as the maximum values of correction among all particles.
We mainly use PCs(Pentium III) equipped with MDGRAPE-2,
which accelerates calculations of momentum-independent forces
including the long-range Coulomb force.

\begin{figure*}
\resizebox{15cm}{!}{\includegraphics{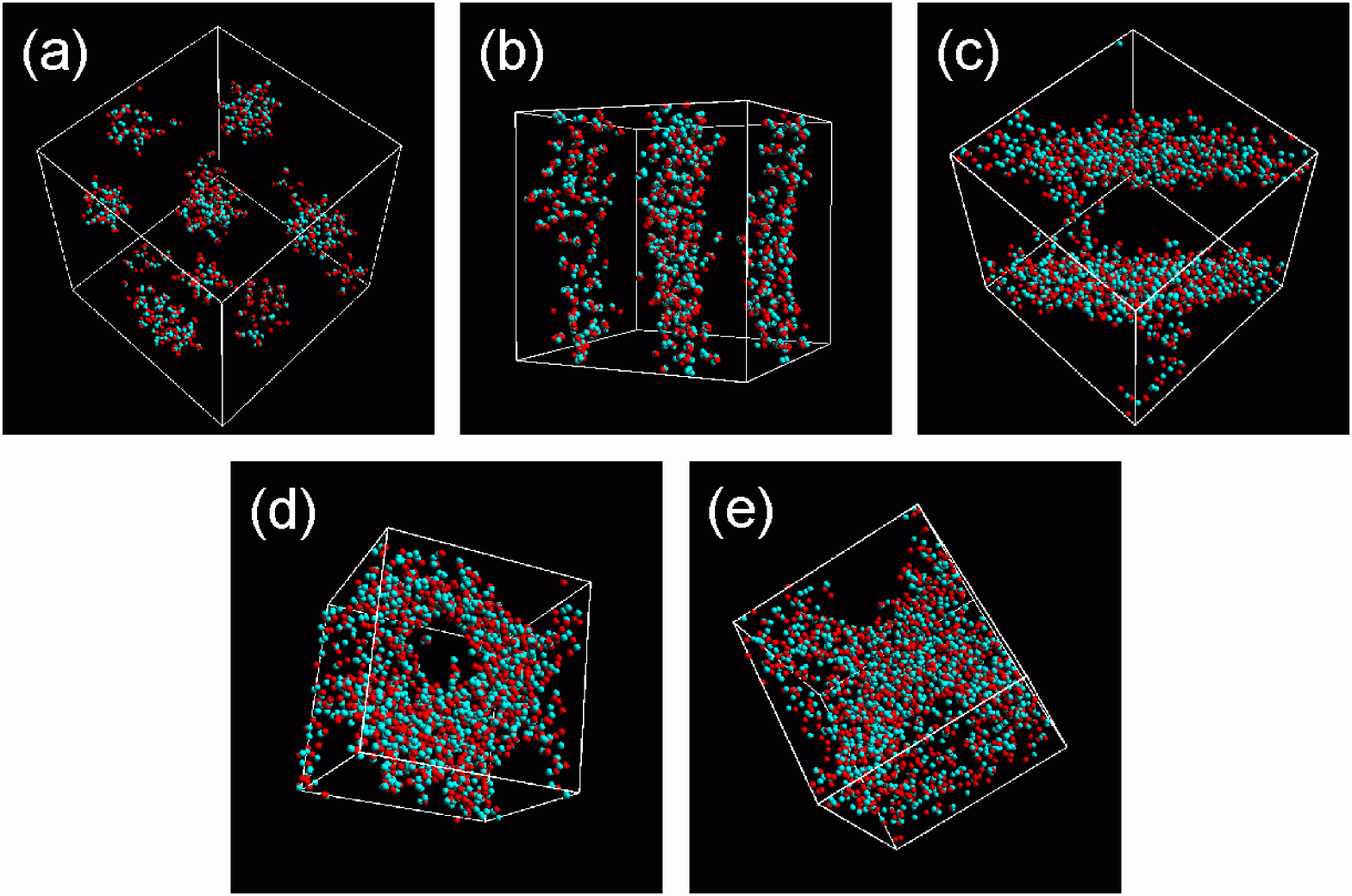}}
\caption{\label{fig pasta sym}(Color)\quad
  The nucleon distributions of typical phases with simple structures
  of cold matter at $x=0.5$;
  (a) sphere phase, $0.1 \rho_{0}$ ($L_{\rm box}=43.65\ {\rm fm}$, $N=1372$);
  (b) cylinder phase, $0.225 \rho_{0}$ ($L_{\rm box}=38.07\ {\rm fm}$, $N=2048$);
  (c) slab phase, $0.4 \rho_{0}$ ($L_{\rm box}=31.42\ {\rm fm}$, $N=2048$);
  (d) cylindrical hole phase, $0.5 \rho_{0}$ ($L_{\rm box}=29.17\ {\rm fm}$, $N=2048$) and
  (e) spherical hole phase, $0.6 \rho_{0}$ ($L_{\rm box}=27.45\ {\rm fm}$, $N=2048$),
  where $L_{\rm box}$ is the box size.
  The red particles represent protons and the green ones represent neutrons.
  }
%\end{figure*}
\vspace{0.5cm}
%\begin{figure*}
\resizebox{15cm}{!}{\includegraphics{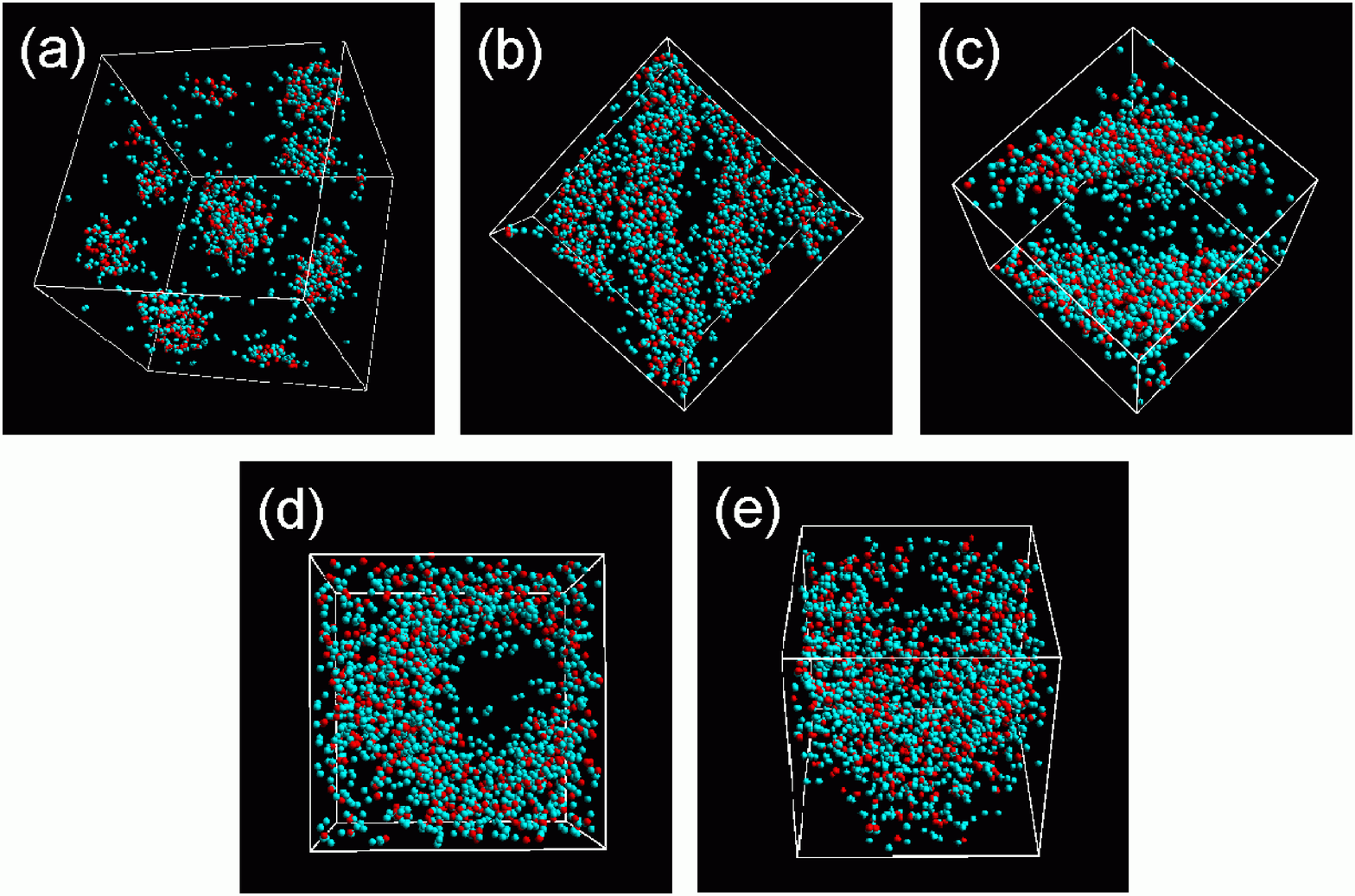}}
\caption{\label{fig pasta x0.3}(Color)\quad
  Same as Fig.\ \ref{fig pasta sym} at $x=0.3$;
  (a) sphere phase, $0.1 \rho_{0}$ ($L_{\rm box}=49.88\ {\rm fm}$, $N=2048$);
  (b) cylinder phase, $0.18 \rho_{0}$ ($L_{\rm box}=41.01\ {\rm fm}$, $N=2048$);
  (c) slab phase, $0.35 \rho_{0}$ ($L_{\rm box}=32.85\ {\rm fm}$, $N=2048$);
  (d) cylindrical hole phase, $0.5 \rho_{0}$ ($L_{\rm box}=29.17\ {\rm fm}$, $N=2048$) and
  (e) spherical hole phase, $0.55 \rho_{0}$ ($L_{\rm box}=28.26\ {\rm fm}$, $N=2048$).
  The red particles represent protons and the green ones represent neutrons.
  }
\end{figure*}

Shown in Figs.\ \ref{fig pasta sym} and \ref{fig pasta x0.3}
are the resultant nucleon distributions
of cold matter at $x=$ 0.5 and 0.3, respectively.
We can see from these figures that
the phases with rodlike and slablike nuclei,
cylindrical and spherical bubbles,
in addition to the phase with spherical nuclei are reproduced
in both the cases of $x=$ 0.5 and 0.3.
We here would like to mention the reasons of discrepancies
between the present result and the result obtained by Maruyama et al.
which says ``the nuclear shape may not have these simple symmetries''
\cite{maruyama}.
One of the most crucial reasons seems to be the difference in
treatment of the Coulomb interaction.
In the present simulation,
we calculate the long range Coulomb interaction in a consistent way
using the Ewald method.
For the system of interest where the Thomas-Fermi screening length is
comparable to or larger than the size of nuclei,
this treatment is more adequate than that which inctroduces
an artificial cutoff distance as in Ref. \cite{maruyama}.
The other crucial reason is the difference
in the relaxation time scales $\tau$ fm/$c$;
we set $\tau \sim O(10^{3}-10^{4})$ in the present work,
but Maruyama et al. set $\tau \sim$ several$\times 10^{3}$ fm/$c$
\cite{maruyama comm}.
In our simulation, we can reproduce the bubble-phases
[see (d) and (e) of Figs.\ \ref{fig pasta sym} and \ref{fig pasta x0.3}]
with $\tau \sim 10^{3}$ fm/$c$
and the nucleus-phases
[see (b) and (c) of Figs.\ \ref{fig pasta sym} and \ref{fig pasta x0.3}]
with $\tau \sim O(10^{4})$ fm/$c$.
However, the matter in the density region
corresponding to a nucleus-phase
is quenched in an amorphous-like state when $\tau \alt 10^{3}$ fm/$c$.
In the present work, we take $\tau$ much larger than
typical time scale $\tau_{\rm th} \sim O(100)$ fm/$c$
for nucleons to thermally diffuse in
the distance of $l \sim 10$ fm at $\rho \simeq \rho_{0}$ and
$k_{\rm B}T \simeq 1$ MeV.
This temperature is lower than the typical value of
the liquid-gas phase transition temperature
in the density region of interest,
it is thus considered that our results are thermally relaxed
in a satisfying level.

\begin{figure}
\rotatebox{270}{
\resizebox{5cm}{!}
{\includegraphics{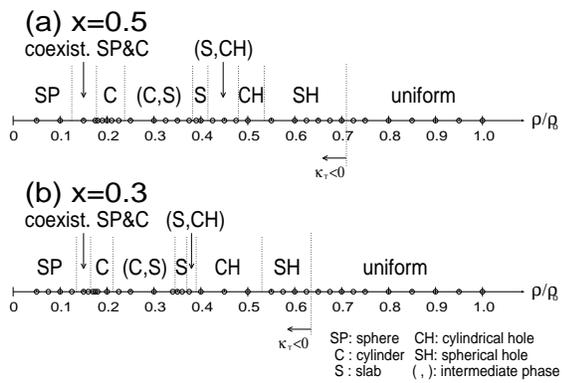}}}
\caption{\label{phase diagram 1}
  Phase diagrams of cold matter at $x=0.5$ (a) and $x=0.3$ (b).
  Matter is unstable against phase separation in the density region
  shown as $\kappa_{\rm T} < 0$,
  where $\kappa_{\rm T}$ is the isothermal compressibility.
  The symbols SP, C, S, CH and SH stand for nuclear shapes,
  i.e., sphere, cylinder, slab, cylindrical hole and spherical hole,
  respectively.
  The parentheses (A,B) show intermediate phases between A-phase and B-phase
  suggested in this work.
  They have complicated structures different from those of
  both A-phase and B-phase.
  Simulations have been carried out at densities denoted by small circles.}
\end{figure}

Phase diagrams of matter in the ground state are shown in
Figs.\ \ref{phase diagram 1}(a) and (b)
for $x=$ 0.5 and 0.3, respectively.
As can be seen from these figures,
the obtained phase diagrams basically reproduce
the sequence of the energetically favored nuclear shapes
predicted by simple discussions \cite{hashimoto}
which only take account of the Coulomb and surface effects;
this prediction is that the nuclear shape changes like
\ sphere $\rightarrow$ cylinder $\rightarrow$ slab $\rightarrow$
cylindrical hole $\rightarrow$ spherical hole $\rightarrow$ uniform,
with increasing density.
Comparing Figs.\ 3 (a) and (b),
we can see that the phase diagram shifts towards the lower density side
with decreasing $x$, which is due to the tendency
that the saturation density is lowered as the neutron excess increases.
It is remarkable that the density dependence of the nuclear shape,
except for spherical nuclei and bubbles, is quite sensitive
and phases with intermediate nuclear shapes
which are not simple as shown in Figs. 1 and 2
are observed in two density regions:
one is between the cylinder phase and the slab phase,
the other is between the slab phase and the cylindrical hole phase.
We note that these phases are different from coexistence phases
with nuclei of simple shapes, which will be referred to as
``intermediate phases''. %in the rest of this paper.

\subsection{QMD Simulations for $x=0.1$}

We have also performed QMD simulations of matter with proton fraction $x=0.1$
as a more realistic condition for the neutron star matter.
In this case, we have to deal with a larger system than in the cases of
$x=0.5$ and 0.3 because enough number of protons
for reproducing several nuclei in a simulation box
are required to obtain significant results;
protons play an important role in generating the long-range order
due to their electric charge.
We have investigated the neutron-rich matter at $x=0.1$ with 10976 nucleons,
in which 1098 protons and 9878 neutrons are contained.
Following basically the same procedure that used for
the cases of $x=0.5$ and 0.3
(see Section \ref{sim p-rich} for detail),
we tried to obtain the ground-state matter.
However, in the present case, we quickly relax from the initial state
at $k_{\rm B}T \sim 20$ MeV to the state at $k_{\rm B}T \sim 10$ MeV,
at which matter is still uniform, with a Nos\'e-Hoover-like thermostat
\cite{nose,hoover}
which will be discussed in another paper \cite{qmd hot}.
After the relaxation at $k_{\rm B}T \sim 10$ MeV
for $\sim 4000$ -- 7000 fm/$c$, we then cool down the system
with a relaxation time scale $\tau \sim O(10^{4})$ fm/$c$
using the QMD equations of motion with friction terms (\ref{qmdeom fric}).
These simulations are performed by Fujitsu VPP 5000 equipped with MDGRAPE-2.

\begin{figure}[htbp]
\begin{center}
\rotatebox{0}{
\resizebox{6.5cm}{!}
{\includegraphics{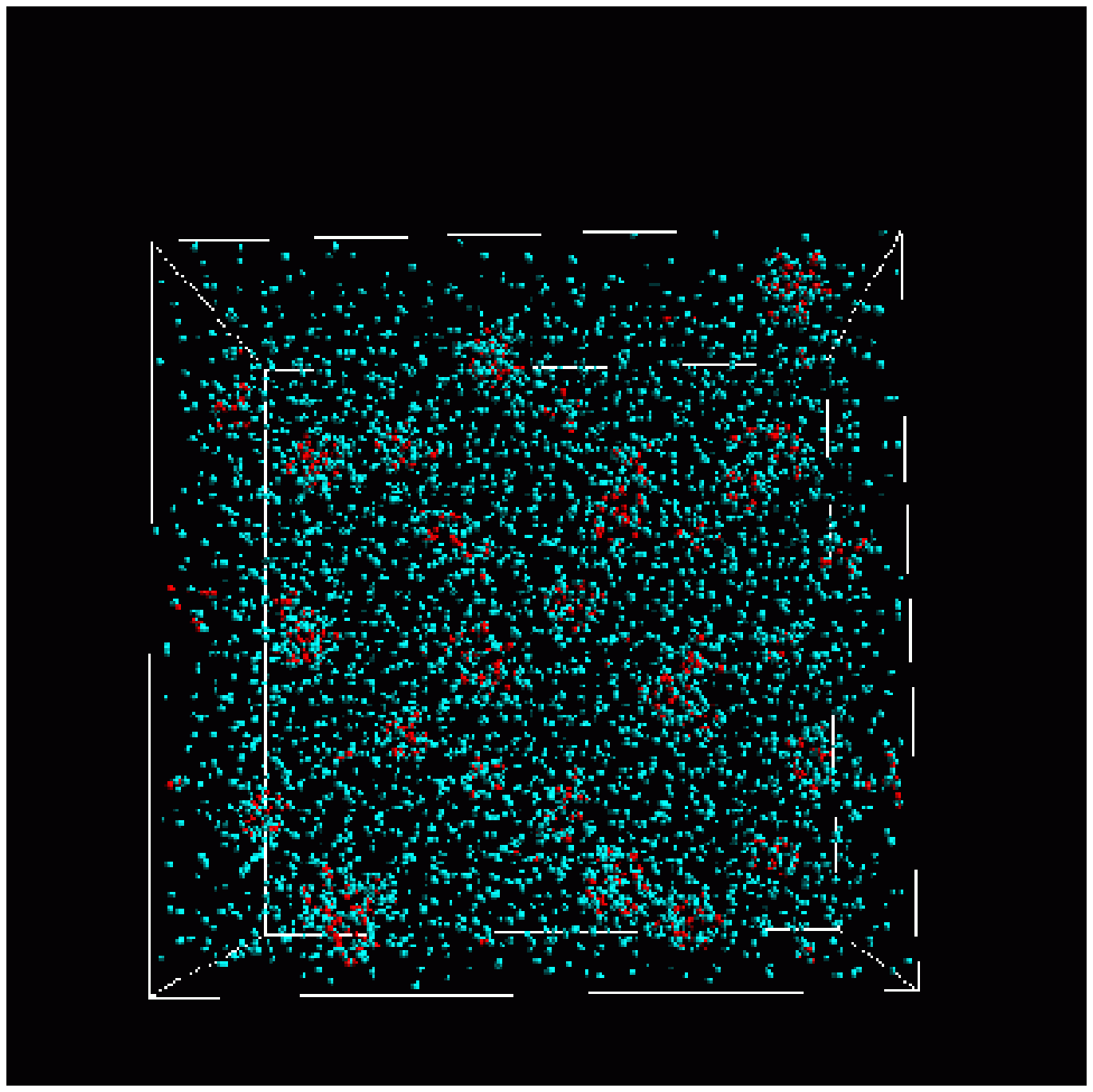}}}
\caption{\label{fig sphere x0.1}(Color)\quad
  The nucleon distribution of sphere phase in cold matter at $x=0.1$.
  The nucleon density $\rho$ and the size $L_{\rm box}$ of the simulation box
  are $\rho = 0.075 \rho_{0}$ and $L_{\rm box}=96.08$ fm.
  The red particles represent protons and the green ones represent neutrons.
  }
\vspace{2cm}
\rotatebox{0}{
\resizebox{6.5cm}{!}
{\includegraphics{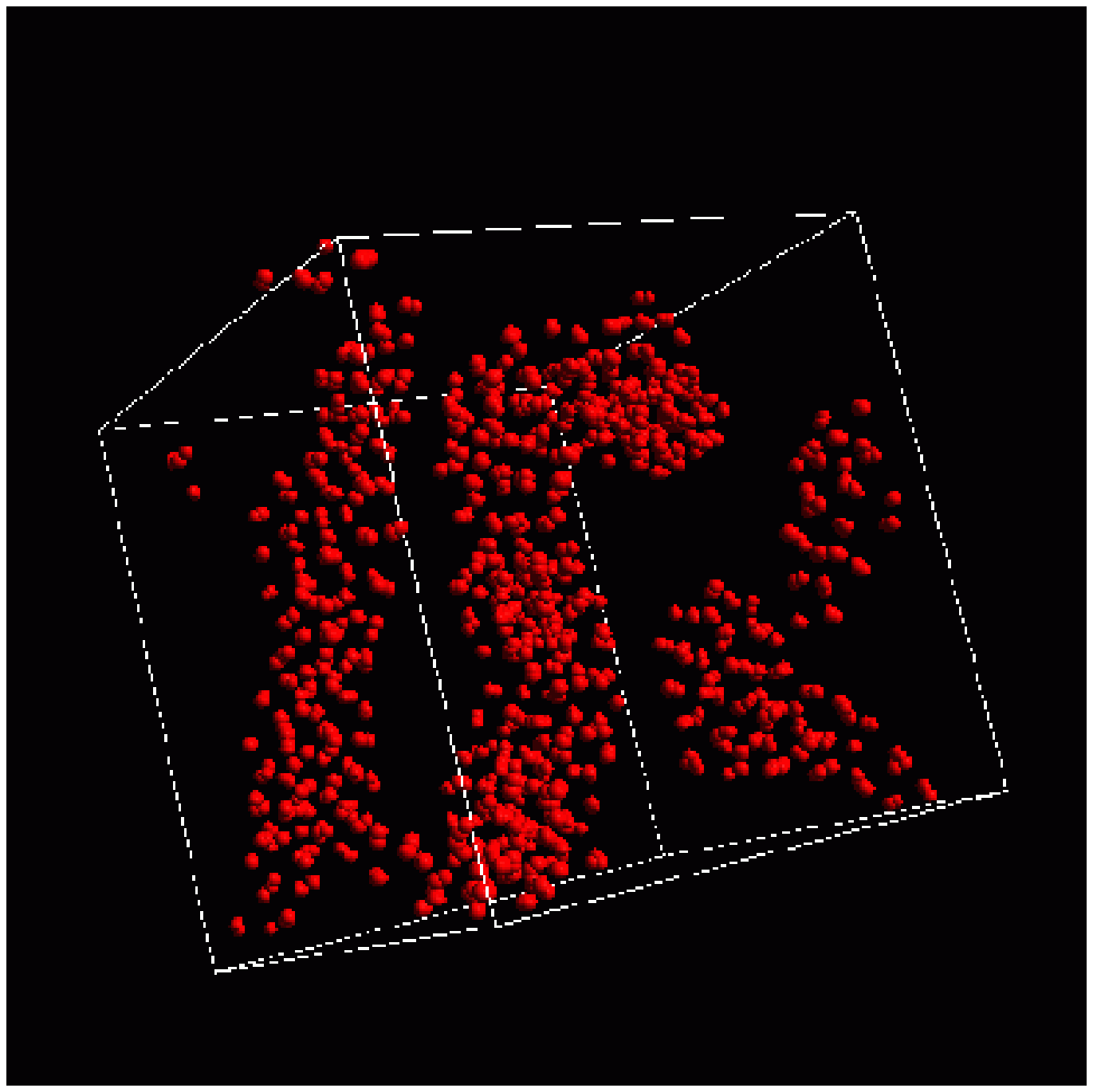}}}
\caption{\label{fig cylinder x0.1}(Color)\quad
  The proton distribution of cylinder phase in cold matter at $x=0.1$.
  The nucleon density $\rho$ and the size $L_{\rm box}$ of the simulation box
  are $\rho = 0.2 \rho_{0}$ and $L_{\rm box}=69.29$ fm.
  Neutrons which spread over the whole space are not depicted in this figure.
  }
\end{center}
\end{figure}

Some resultant nucleon distributions are shown
in Figs.\ \ref{fig sphere x0.1} and \ref{fig cylinder x0.1},
which correspond to the sphere phase and the cylinder phase, respectively.
As can be seen in Fig.\ \ref{fig sphere x0.1},
dripped neutrons spread over the whole region in the simulation box,
which lead to smaller density contrast compared with that for
the cases of $x=0.5$ and 0.3 depicted in Figs.\ \ref{fig pasta sym} and
\ref{fig pasta x0.3}, respectively.

\begin{figure}[htbp]
\begin{center}
\rotatebox{270}{
\resizebox{3cm}{!}
{\includegraphics{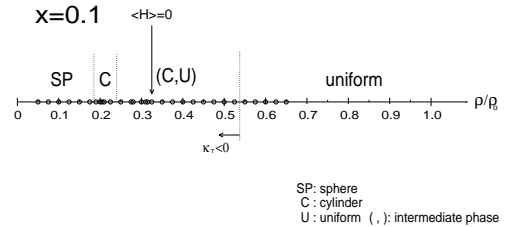}}}
\caption{\label{phase diagram 2}
  Phase diagram of cold matter at $x=0.1$.
  Matter is unstable against phase separation in the density region
  shown as $\kappa_{\rm T} < 0$,
  where $\kappa_{\rm T}$ is the isothermal compressibility.
  The symbols SP, C and U stand for shapes of nuclear matter region,
  i.e., sphere, cylinder and uniform, respectively.
  The density at which the area-averaged mean curvature of nuclear surface
  is zero is denoted by $\langle H\rangle =0$.
  However, slab phase is not observed in our results even at such a density.
  The parentheses (A,B) show intermediate phases between A-phase and B-phase.
  They have complicated structures different from those of
  both A-phase and B-phase.
  Simulations have been carried out at densities denoted by small circles.}
\end{center}
\end{figure}

The results obtained for $x=0.1$ are summarized in the phase diagram
shown in Fig.\ \ref{phase diagram 2}.
A striking feature is that the wide density region
from $\sim 0.25 \rho_{0}$ to $\sim 0.525 \rho_{0}$
is occupied by an intermediate phase.
The structure of matter seems to change rather continuously from
that consists of branching rodlike nuclei connected to each other
[obtained in the lower density region
of the intermediate phase denoted by (C,U)]
to that consists of branching bubbles connected to each other
[higher density region of the intermediate phase (C,U)].
However, in the present neutron-rich case, ``pasta'' phase with slablike nuclei
cannot be obtained as far as we have investigated,
which will be discussed at the end of the next section.
It is also noted that the density at which matter turns into uniform
is lower than those in the cases of $x=0.5$ and 0.3,
which is consistent with the tendency
that matter becomes the more neutron-rich,
the saturation density becomes the lower.

%%%%%%%%%%%%%%%%%%%%%%%%%%%%%%%%%%%%%%%%
\section{Analysis of the Structure of Matter}

\subsection{Two-point correlation functions\label{2pcf}}

To analyze the spatial distribution of nucleons,
we calculate two-point correlation function $\xi_{ii}$
for nucleon density field $\rho^{(i)}$ $(i=N,p,n;$ where $N$ stands for
nucleons$)$.
$\xi_{ii}$ is here defined as
\begin{eqnarray}
  \xi_{ii}&=&\frac{1}{4\pi} \int d\Omega_{\bf r}\ \frac{1}{V} \int d^{3}{\bf x}\
    \delta_{i}({\bf x}) \delta_{i}({\bf x+r})\\
    &\equiv& \langle \delta_{i}({\bf x}) \delta_{i}({\bf x+r})
    \rangle_{{\bf x}, \Omega_{\bf r}}\ ,
\end{eqnarray}
where $\langle\cdots\rangle_{{\bf x}, \Omega_{\bf r}}$ denotes
an average over the position ${\bf x}$ and the direction of ${\bf r}$,
and $\delta_{i}({\bf x})$ is the fluctuation of the density field
$\rho^{(i)}({\bf x})$ given by
\begin{equation}
  \delta_{i}({\bf x}) \equiv \frac{\rho^{(i)}({\bf x}) - \overline{{\rho}^{(i)}}}
  {\overline{\rho^{(i)}}}\ ,
\end{equation}
with
\begin{equation}
  {\overline{\rho^{(i)}}} \equiv \frac{N_{i}}{V}\ .
\end{equation}

We construct the nucleon density distribution $\rho^{(i)}({\bf x})$
from a data set of the centers of position of the nucleons
by the following procedure.
We first set $64^{3}$ (for $x=0.5$ and 0.3) or $128^{3}$
(for $x=0.1$) grid points in the simulation box
and then distribute particle numbers on each grid point using
the cloud-in-cell method (see, e.g., Ref.\ \cite{hockney}).
Next, we carry out the smoothing procedure
in the discrete Fourier space with a Gaussian smoothing function
corresponding to the distribution of the wave packet
given by Eq.\ (\ref{packet}).
The density distributions
$\rho^{(i)}({\bf x})=\sum_{k=1}^{N} n^{(i)}_{k} |\phi_{k}({\bf x})|^{2}$,
where $n^{(i)}_{k}=0$ or 1 projects on particle type $i$,
%%%%%
%$\rho^{(i)}({\bf x}) = |\Phi_{i}({\bf x})|^{2}$,
%where $\Phi_{i}({\bf x})$ is the QMD trial wave function of nucleons
%given by Eq.\ (\ref{trial func}),
%%%%%
in the discrete real space can be obtained
by the inverse Fourier transformation.
The Fourier transformations are performed using the FFT algorithm.

\begin{figure}
\resizebox{8cm}{!}
{\includegraphics{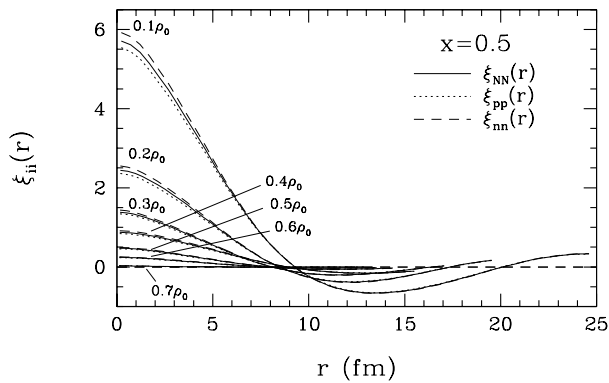}}
\caption{\label{corr x0.5}
  Two-point correlation functions of density fluctuations
  calculated for $x=0.5$.  The solid lines show the two-point
  correlation function for nucleon density distributions;
  the dotted lines, that for proton density distributions;
  the dashed lines, that for neutron density distributions.
  }
\end{figure}

\begin{figure}
\resizebox{8cm}{!}
{\includegraphics{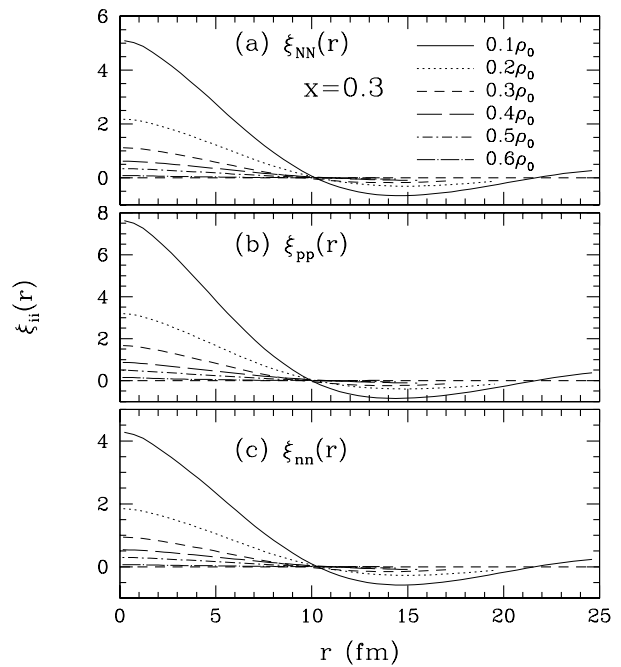}}
\caption{\label{corr x0.3}
  Two-point correlation functions of density fluctuations
  calculated for $x=0.3$; (a) for nucleon density distributions;
  (b) for proton density distributions;
  (c) for neutron density distributions.
  }
\end{figure}

\begin{figure}
\resizebox{8cm}{!}
{\includegraphics{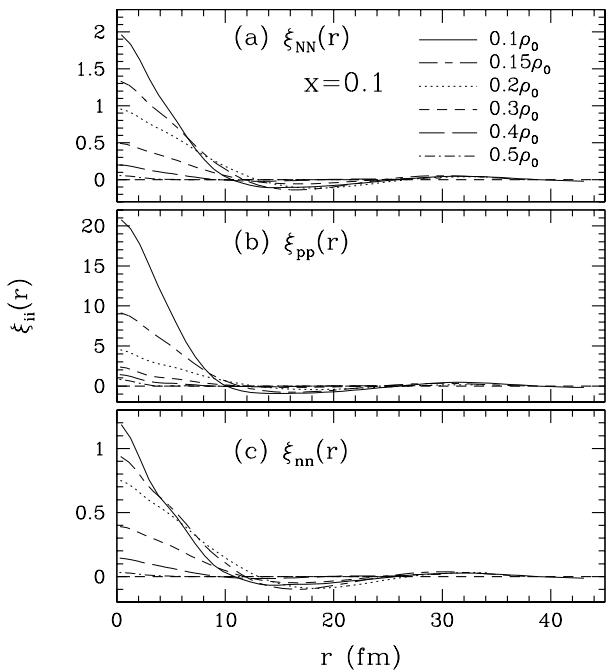}}
\caption{\label{corr x0.1}
  The same as Fig.\ \ref{corr x0.3} for $x=0.1$.
  }
\end{figure}

The resultant two-point correlation functions $\xi_{NN}(r)$, $\xi_{pp}(r)$
and $\xi_{nn}(r)$ at various densities below $\rho_{\rm m}$ at which
matter becomes uniform at zero temperature
for $x=0.5$, 0.3 and 0.1 are plotted in
Figs.\ \ref{corr x0.5}, \ref{corr x0.3} and \ref{corr x0.1}, respectively.
We can see the general tendency, which is common
for the different values of $x$, that the amplitude of $\xi_{ii}(r)$
decreases with increasing  the density.
It is noted that even though the change in the amplitude of $\xi_{ii}(r)$
is quite noticeable, the smallest zero-point $r=r_{0}$ of $\xi_{ii}(r)$
takes similar values at various densities especially for $x=0.5$ and 0.3.
This feature means that the typical length scales of the nuclear structures,
i.e., the internuclear distance and the nuclear radius, keep comparable
at subnuclear densities from $\sim 0.1\rho_{0}$ to $\rho_{\rm m}$,
which is consistent with the results obtained by the previous works
(see, e.g., Refs.\ \cite{lorenz,oyamatsu,gentaro1,gentaro2}).
This behavior just below $\rho_{\rm m}$ will be discussed further
concerning a problem about the properties of the transition to uniform matter.

We also note that a strong attractive force
acting between a proton and a neutron leads to the good agreement
of the zero-points of $\xi_{pp}$ and $\xi_{nn}$ even for $x=0.3$ and 0.1
as well as for $x=0.5$ although the zero-point $r_{0}$ of $\xi_{nn}$ is
$\sim 0.3$ fm ($\alt 0.5$ fm) larger than that of $\xi_{pp}$ for $x=0.3$ (0.1)
at each density.
This shows that the phase of the density fluctuations of protons
and neutrons correlate so strongly with each other at zero temperature
that they almost coincide.

As can be seen by comparing $\xi_{NN}(r)$ for different values of
the proton fraction $x$,
the amplitude decreases and the value of the smallest zero-point $r_{0}$
increases with decreasing $x$.  This behavior means that,
as matter becomes more neutron-rich, not only the nucleon density distribution
gets smoother but also the spatial structure becomes larger.

Let us then examine $\xi_{ii}$ for each value of the proton fraction.
For symmetric matter ($x=0.5$), protons and neutrons are equivalent
except for the mass difference and the Coulomb interaction.
Therefore, the two-point correlation functions $\xi_{NN}$, $\xi_{pp}$
and $\xi_{nn}$ are almost the same at larger values of $r\agt r_{0}$.
At smaller values of $r$, $\xi_{pp}$ is slightly smaller than $\xi_{nn}$
because the repulsive Coulomb interaction among protons tends to reduce
the proton density inhomogeneity especially in the smaller scale.
For asymmetric matter ($x=0.3$ and 0.1), on the contrary,
the amplitude of $\xi_{nn}$ is much smaller than that of $\xi_{pp}$
due to the dripped neutrons which distribute rather uniformly
outside the nuclei.

\subsection{Transition to uniform matter}

\begin{figure}
\resizebox{8cm}{!}{\includegraphics{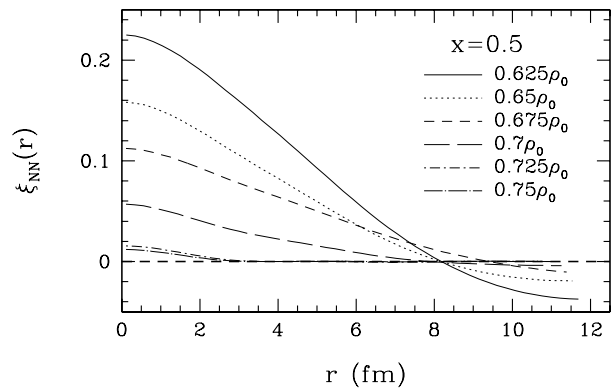}}
\caption{\label{corr2 melt}
  Two-point correlation function $\xi_{NN}(r)$ of the nucleon density
  fluctuation.  $\xi_{NN}(r)$ is calculated for $x=0.5$ and
  for densities around $\rho_{\rm m}$ from 0.625$\rho_{0}$ to 0.75$\rho_{0}$.
  }
\end{figure}

Let us here examine the properties of the transition
from the phase with spherical bubbles to uniform matter for $x=0.5$.
For this purpose, two-point correlation function
of the nucleon density fluctuation is useful.
In Fig.\ \ref{corr2 melt}, we thus plot the two-point correlation function
of the nucleon density fluctuation $\xi_{NN}(r)$
for several densities around the melting density $\rho_{\rm m}$.
To compute $\xi_{NN}(r)$, we use a 1372-nucleon system cooled down
until the temperature gets $\sim$ 0.05 MeV
by QMD equations of motion (\ref{qmdeom fric}) with friction terms.

For uniform phase, $\xi_{NN}(r)$ should be zero
except for the contribution of short-range correlation.
The behavior of $\xi_{NN}(r)$ shows that $\rho_{\rm m}$ lies between
0.7$\rho_{0}$ and $0.725\rho_{0}$ at $x=0.5$
[see also Fig.\ \ref{phase diagram 1}(a)]
above which long-range correlation disappears.
It is noted that the smallest value of $r=r_{0}$ at which
$\xi_{NN}(r)=0$ keeps around 8 fm even at densities
just below $\rho_{\rm m}$.
This means that, for $x=0.5$, the phase with spherical bubbles
whose radii are around $r_{0}$ suddenly disappear rather than they shrink
gradually and the system turns into uniform with increasing the density
because the quantity $r_{0}$ nearly corresponds to the half wave-length
of the inhomogeneous density profile.
The discontinuous change in the density profile indicates that
the transition between the phase with spherical bubbles and the uniform
phase is of first order.
This conclusion is also obtained in the previous calculations for which
the spatial structure of the nuclear matter region and/or
the shape of the density profile are assumed
\cite{oyamatsu,lorenz,ogasawara,gentaro1,gentaro2}.
In the present work, we have confirmed the first order nature
by QMD simulations without these assumptions
as several authors have done so without the assumptions
by the Thomas-Fermi approximation
\cite{williams,lassaut}.

For the cases of $x=0.3$ and 0.1, we could not see the significant
sign of the first order nature of the transition
between the mixed phase and the uniform phase
because the amplitude of $\xi_{NN}(r)$ is quite small
just below $\rho_{\rm m}$.
Further study is necessary to determine the properties of the transition
for theses cases of asymmetric matter.

\subsection{Minkowski functionals}

To extract the morphological characteristics of the nuclear shape changes
and the intermediate phases,
we introduce the Minkowski functionals
(see, e.g., Ref. \cite{minkowski} and references therein;
a concise review is provided by Ref.\ \cite{jens})
as geometrical and topological measures of the nuclear surface.
Let us consider a homogeneous body $K \in {\cal R}$
in the $d$-dimensional Eucledian space,
where ${\cal R}$ is the class of such bodies.
Morphological measures are defined as functionals
$\varphi: {\cal R} \rightarrow {\bf R}$
which satisfy the following three general properties:
\begin{itemize}
  \item[(1)] {\it Motion invariance} :\quad
    The functional is independent of the position and the direction
    of the body, i.e.,
    \begin{equation}
      \varphi (K) = \varphi (gK)\ ,
    \end{equation}
    where $g$ denotes any translations and rotations.
  \item[(2)] {\it Additivity} :\quad
    The functional of the union of two bodies should behave like a volume.
    The contribution of the overlapping region should be subtracted
    , i.e.,
    \begin{equation}
      \varphi (K_{1} \cup K_{2})
      = \varphi(K_{1})+\varphi(K_{2})-\varphi(K_{1} \cap K_{2})\ ,
    \end{equation}
    where $K_{1},K_{2} \in {\cal R}$.
  \item[(3)] {\it Continuity} :\quad
    If the body is approximated with pixels,
    the functional of the approximate body
    converges to that of the original body when the pixels get smaller
    , i.e.,
    \begin{equation}
      \lim_{n\to\infty} \varphi(K_{n}) = \varphi(K)
      \quad {\rm as} \quad \lim_{n\to\infty} K_{n}=K\ ,
    \end{equation}
      where $K$ is a convex body and
      $\{ K_{n} \}$ is a sequence of convex bodies.
\end{itemize}
Hadwiger's theorem in integral geometry states that
there are just $d+1$ independent functionals
which satisfy the above properties;
they are known as Minkowski functionals.
In three dimensional space, four Minkowski functionals are related to
the volume, the surface area, the integral mean curvature
and the Euler characteristic.

In classifying nuclear shapes including those of the intermediate phases
obtained in our simulations, the integral mean curvature
and the Euler characteristic are useful, which will be discussed later.
Both are described by surface integrals of the following local quantities:
the mean curvature $H = (\kappa_{1}+\kappa_{2})/2$ and
the Gaussian curvature $G = \kappa_{1} \kappa_{2}$,
i.e., $\int_{\partial K} H dA$ and
$\chi \equiv \frac{1}{2 \pi} \int_{\partial K} G dA$,
where $\kappa_{1}$ and $\kappa_{2}$ are the principal curvatures and
$dA$ is the area element of the surface of the body $K$.
The Euler characteristic $\chi$ is a purely topological quantity
and is given by
\begin{eqnarray}
  \chi & = & \mbox{(number of isolated regions)}
  - \mbox{(number of tunnels)} \nonumber\\
  && + \mbox{(number of cavities)}.
  \label{euler}
\end{eqnarray}
Thus $\chi > 0$ for the sphere and the spherical hole phases
and the coexistence phase of spheres and cylinders,
and $\chi = 0$ for the other ideal ``pasta'' phases,
i.e. the cylinder, the slab and the cylindrical hole phases
which consist of infinitely long rods, infinitely extending slabs and
infinitely long cylindrical holes, respectively.
We introduce the area-averaged mean curvature,
$\langle H \rangle \equiv \frac{1}{A}\int H dA$,
and the Euler characteristic density, $\chi / V$, as normalized quantities,
where $V$ is the volume of the whole space.

%%%%%
\subsubsection{Minkowski functionals for $x=0.5$ and 0.3}

We calculate the normalized Minkowski functionals,
i.e., the volume fraction $u$, the surface area density $A/V$,
the area-averaged mean curvature $\langle H \rangle$ and
the Euler characteristic density $\chi/V$
for $x=0.5$ and 0.3 by the following procedure.
As described in Subsection \ref{2pcf},
we first construct proton and nucleon density distributions
$\rho^{(p)}({\bf r})=\sum_{k=1}^{N} n^{(p)}_{k} |\phi_{k}({\bf r})|^{2}$ and
$\rho({\bf x})=\sum_{k=1}^{N} |\phi_{k}({\bf x})|^{2}$,
where $n^{(p)}_{k}=0$ or 1 is the isospin projection on the proton state.
%%%%%
%We first construct proton and nucleon density distributions
%$\rho^{(p)} ({\bf r}) = \left| \Phi_{p} ({\bf r}) \right|^{2}$ and
%$\rho ({\bf r}) = \left| \Phi ({\bf r}) \right|^{2}$
%as described in Subsection \ref{2pcf}.
%%%%%
We set a threshold proton density $\rho_{p, \rm{th}}$ and then calculate
$f(\rho_{p, \rm{th}}) \equiv V(\rho_{p, \rm{th}})/A(\rho_{p, \rm{th}})$,
where $V(\rho_{p, \rm{th}})$ and $A(\rho_{p, \rm{th}})$ are
the volume and the surface area of the regions
in which $\rho^{(p)}({\bf r}) \geq \rho_{p, \rm{th}}$.
We find out the value $\rho_{p, \rm{th}} = \rho_{p, \rm{th}}^{*}$
where $\frac{d^{2}}{d \rho_{p, \rm{th}}^{2}} f(\rho_{p, \rm{th}}^{*}) = 0$
and define the regions in which
$\rho^{(p)}({\bf r}) \geq \rho_{p, \rm{th}}^{*}$ as nuclear regions.
For spherical nuclei, for example, $\rho_{p, \rm{th}}^{*}$
corresponds to a point of inflection of a radial density distribution.
In the most phase-separating region,
the values of $\rho_{\rm th}^{*}$ distribute
in the range of about $0.07-0.09$ fm$^{-3}$
in the both cases of $x = 0.5$ and 0.3,
where $\rho_{\rm th}^{*}$ is the threshold nucleon density
corresponds to $\rho_{p, \rm{th}}^{*}$.
We then calculate $u$, $A$, $\int H dA$ and $\chi$
for the identified nuclear surface.
We evaluate $u$ by counting the number of pixels at which $\rho^{(p)}({\bf r})$
is higher than $\rho_{p, \rm{th}}^{*}$,
$A$ by the triangle decomposition method,
$\int H dA$ by the algorithm shown in Ref. \cite{minkowski}
in conjunction with a calibration by correction of surface area,
and $\chi$ by the algorithm of Ref. \cite{minkowski} and
by that of counting deficit angles \cite{genus},
which are confirmed that both of them give the same results.

\begin{figure}
\resizebox{8.2cm}{!}
{\includegraphics{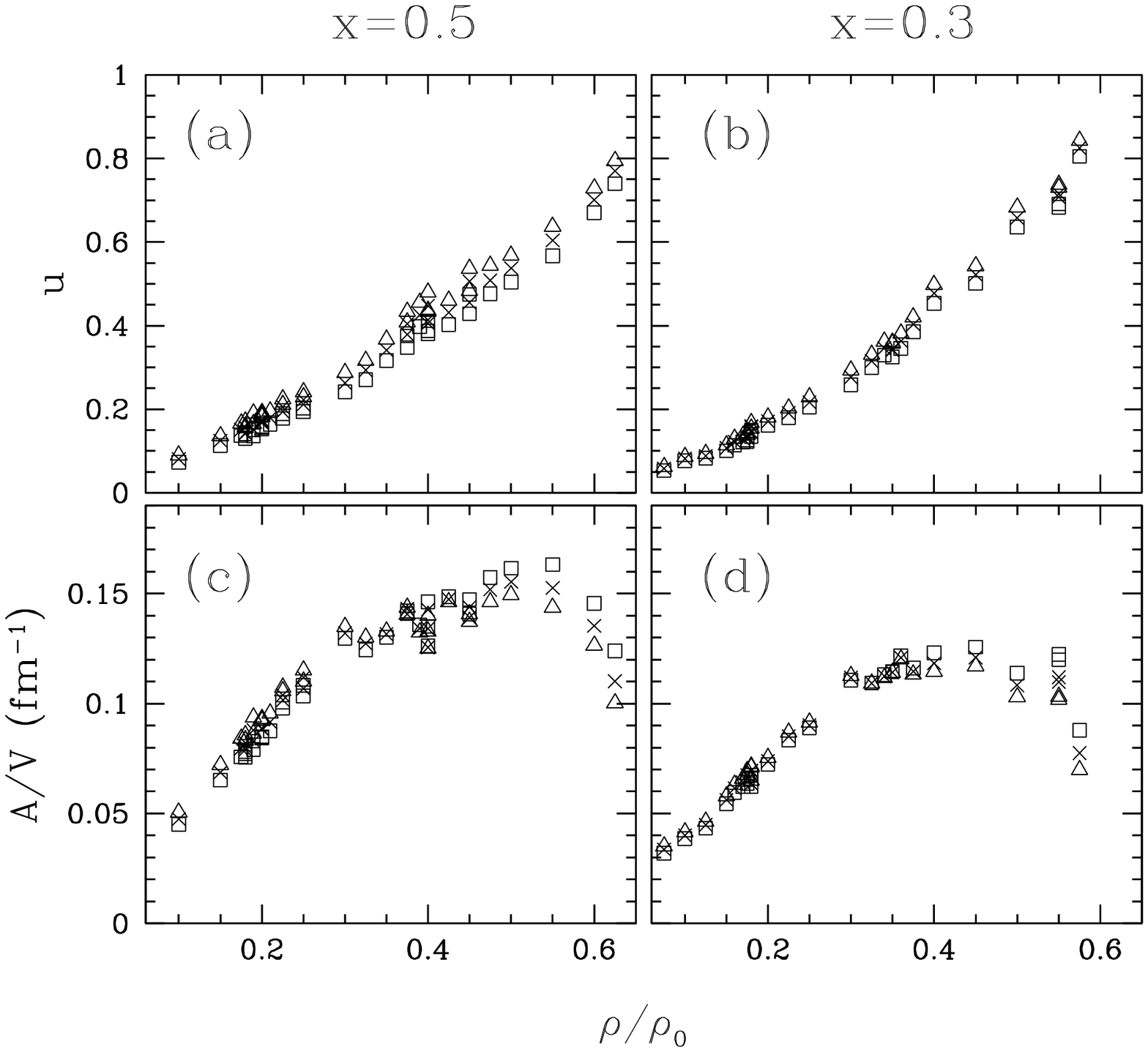}}
\caption{\label{fig minko 1}
  Density dependence of the volume fraction $u$
  and the surface area density $A/V$
  of cold matter at $x=0.5$ (a and c) and
  $x=0.3$ (b and d).
  The crosses show the results for $\rho_{\rm th}=\rho_{\rm th}^{*}$
  and the open triangles and squares show the results for
  $\rho_{\rm th}=\rho_{\rm th}^{*} - 0.05\rho_{0}$ and
  $\rho_{\rm th}^{*} + 0.05\rho_{0}$, respectively.}
%\end{figure}

\vspace{5mm}
%\begin{figure}
\rotatebox{0}{
\resizebox{8.2cm}{!}
{\includegraphics{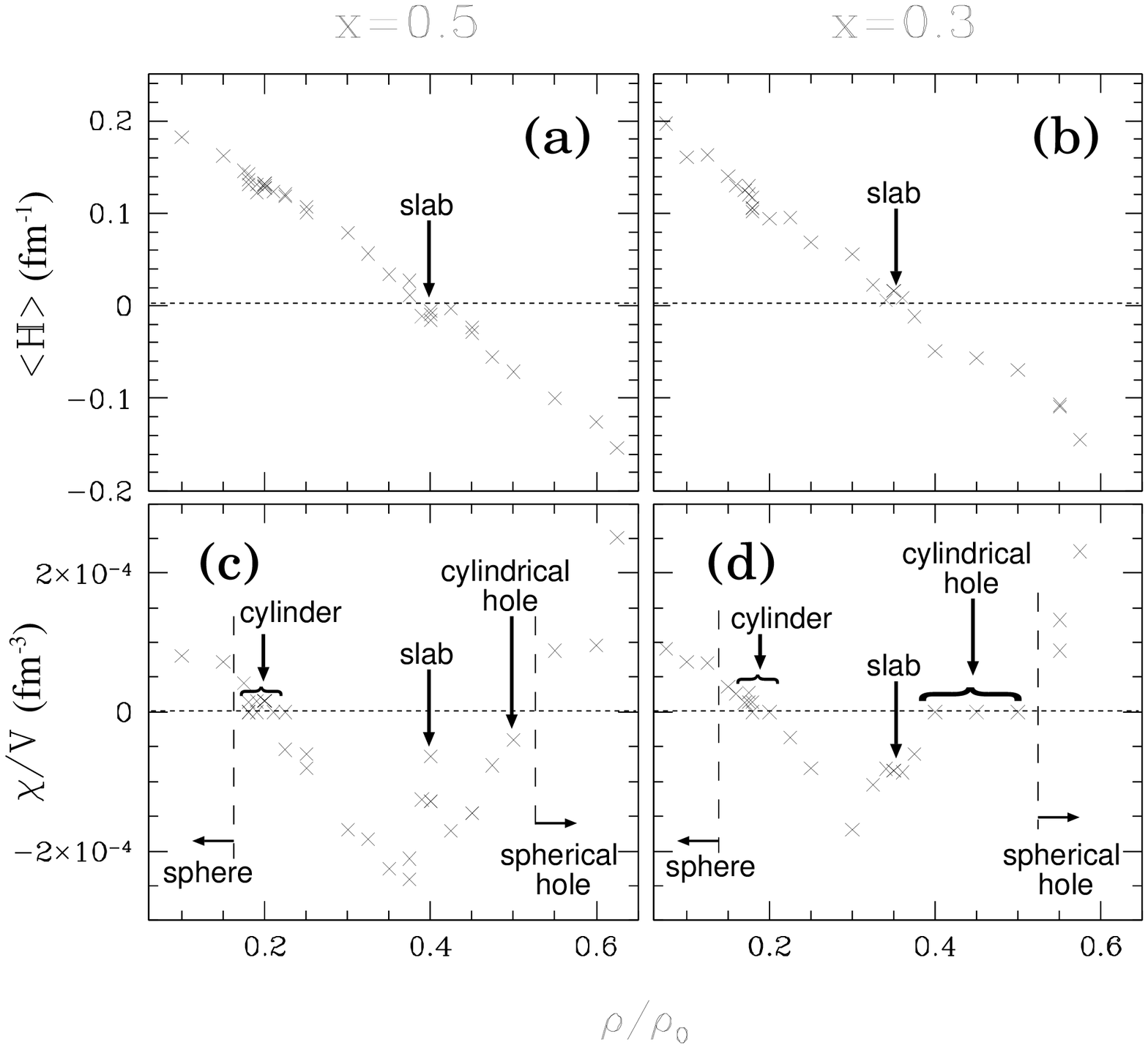}}}
\caption{\label{fig minko 2}
  Density dependence of the area-averaged mean curvature $\langle H\rangle$
  and the Euler characteristic density $\chi/V$
  of cold matter at $x=0.5$ (a and c) and
  $x=0.3$ (b and d).}
\end{figure}

We have plotted the resultant $\rho$ dependence of
$u$, $A/V$, $\langle H \rangle$
and $\chi / V$ for the isodensity surface of
$\rho_{\rm th}=\rho_{\rm th}^{*}$
in Figs. \ref{fig minko 1} and \ref{fig minko 2}.
In addition to the values of $u$, $A/V$ and $\langle H \rangle$
for the isodensity surface of $\rho_{\rm th}=\rho_{\rm th}^{*}$,
we have also investigated those for the isodensity surfaces of
$\rho_{\rm th}=\rho_{\rm th}^{*} \pm 0.05\rho_{0}$
to examine the extent of the uncertainties of these quantities
which stem from the arbitrariness in the definition of the nuclear surface.
As shown in Fig.\ \ref{fig minko 1}, these uncertainties are at most
$\simeq 0.1$ for $u$ and $\simeq 0.25$ fm$^{-1}$ for $A/V$.
For $\langle H \rangle$, we could not observe remarkable differences
from the values for $\rho_{\rm th}=\rho_{\rm th}^{*}$
(they were smaller than 0.015 fm$^{-1}$).
We could not see these kinds of uncertainties in $\chi / V$,
except for the densities near below $\rho_{\rm m}$.

As shown in Fig.\ \ref{fig minko 1}, the volume fraction $u$
of the nuclear regions increases almost monotonically under $\rho_{\rm m}$
in both cases of $x=0.5$ and 0.3.
This feature reflects the incompressible nature of nuclear matter.
It is interesting to see the density dependence of the nuclear surface density
$A/V$ because this quantity is directly related to the surface energy density,
which is one of the key factors in determining the nuclear shape.
Figs.\ \ref{fig minko 1}(c) and \ref{fig minko 1}(d) show that,
as the nucleon density increases, $A/V$ increases at a nearly constant rate
until $\rho \sim 0.3 \rho_{0}$, and then its increasing rate becomes
rather smaller around the density region of the slab phase,
and finally it begins to decrease in the density region of
the cylindrical hole phase or the spherical hole phase.
This general behavior can be understood from the density dependence
of the surface energy density obtained by simple arguments,
which only allow for the nuclear surface and the Coulomb effects
(see, e.g., Refs.\ \cite{hashimoto,gentaro thesis}).

\begin{figure}
\rotatebox{0}{
\resizebox{8.2cm}{!}
{\includegraphics{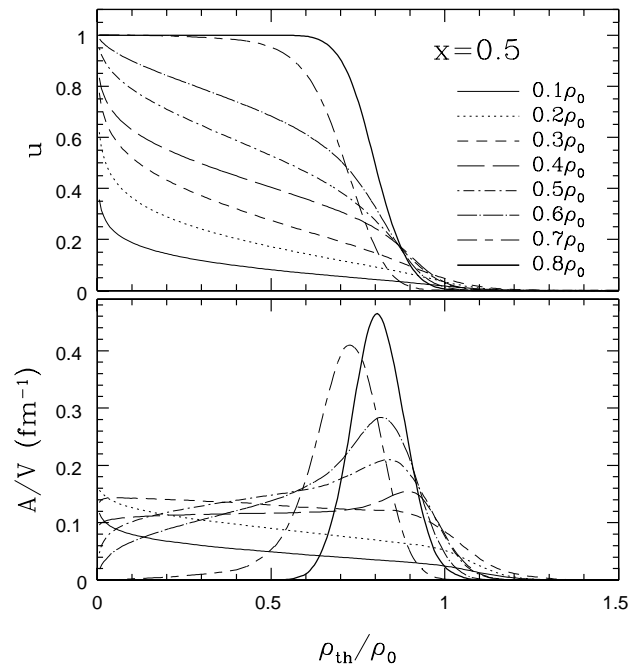}}}
\caption{\label{fig minko thres x0.5}
  Volume fraction $u$ (upper panel) and surface area density $A/V$
  (lower panel) as functions of the threshold density $\rho_{\rm th}$
  calculated for $x=0.5$ and various nucleon densities $\rho$.
  }
\end{figure}

\begin{figure}
\rotatebox{0}{
\resizebox{8.2cm}{!}
{\includegraphics{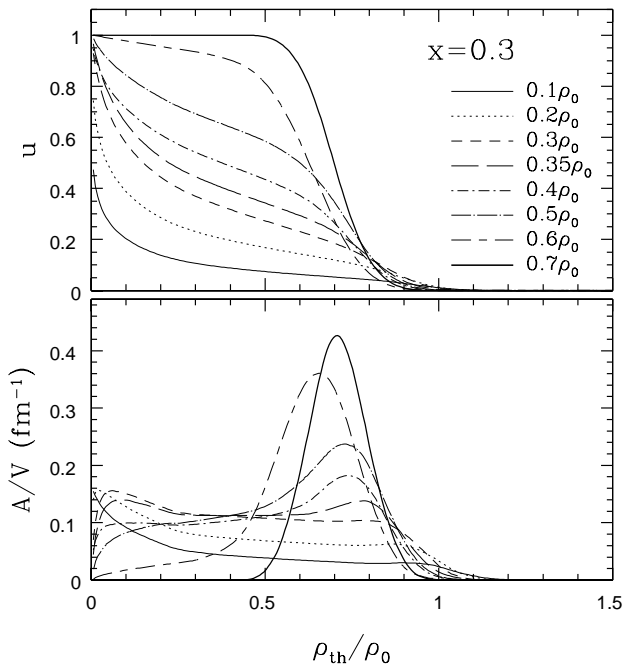}}}
\caption{\label{fig minko thres x0.3}
  The same as Fig.\ \ref{fig minko thres x0.5} for $x=0.3$.
  }
\end{figure}

We also plot $u$ and $A/V$ for the nucleon density distribution $\rho({\bf r})$
as functions of the threshold density $\rho_{\rm th}$
evaluated at various values of $\rho$.
The results for $x=0.5$ and 0.3 are shown in Figs.\ \ref{fig minko thres x0.5}
and \ref{fig minko thres x0.3}, respectively.
Features of nuclear shape changes can be seen in the behavior of
the curves of $A/V$.
Peaks in the higher $\rho_{\rm th}$ region are attributed to
nucleons in the nuclear matter regions
and broad bumps in the lower $\rho_{\rm th}$ region
around 0.05 -- 0.1$\rho_{0}$ observed for $x=0.3$
are due to the dripped neutrons outside nuclei.
$A/V$ in the intermediate $\rho_{\rm th}$ region
in which its slope is nearly constant
mainly comes from contribution of nuclear surfaces.
As the nucleon density $\rho$ increases,
the higher $\rho_{\rm th}$ peak becomes more clear
and the position of the center of the peak finally coincides with $\rho$
in the uniform phase.  This feature shows that the nuclear matter regions
become more uniform with increasing the density.
When the dispersion of the internucleon distance is small,
large surface area caused by the Gaussian density distribution of each nucleon
can be picked up with a single value of $\rho_{\rm th}$.
As can be seen in Fig.\ \ref{fig minko thres x0.3},
the lower $\rho_{\rm th}$ bump, in turn, disappears
with increasing $\rho$.  This is because, as $\rho$ increases,
the dripped neutron gas
becomes more inhomogeneous and tends to distribute close to the nuclear surface
leading to a lower proton fraction in the nuclear matter regions.
It is also noted that, as the nucleon density increases,
the slope in the intermediate $\rho_{\rm th}$ region
changes from negative to positive
at the density corresponding to the phase of slablike nuclei
(0.4$\rho_{0}$ for $x=0.5$ and 0.35$\rho_{0}$ for $x=0.3$),
which is consistent with what is
expected from the sign of $\langle H \rangle$ for the nuclear surface
[see Figs.\ \ref{fig minko 2}(a) and \ref{fig minko 2}(b)].

Let us then focus on $\langle H \rangle$ and $\chi/V$
to classify the nuclear shape.
The behavior of $\langle H \rangle$ shows that
it decreases almost monotonically from positive to negative
with increasing $\rho$ until the matter turns into uniform.
The densities corresponding to $\langle H \rangle \simeq 0$
are about 0.4 and 0.35$\rho_{0}$ for $x=0.5$ and 0.3, respectively;
these values are consistent with the density regions
of the phase with slablike nuclei (see Fig.\ \ref{phase diagram 1}).
As mentioned previously, $\chi / V$ is actually positive
in the density regions corresponding to the phases with spherical nuclei,
coexistence of spherical and cylindrical nuclei, and spherical holes
because of the existence of isolated regions.
As for those corresponding to the phases with cylindrical nuclei,
planar nuclei and cylindrical holes, $\chi / V \simeq 0$.
The fact that the values of $\chi / V$ are not exactly zero
for nucleon distributions shown as the slab phase
in Figs. \ref{fig pasta sym} and \ref{fig pasta x0.3}
reflects the imperfection of these ``slabs'',
which is due to the small nuclear parts connecting the neighboring slabs.
However, we can say that the behaviors of $\chi / V$
plotted in Figs. \ref{fig minko 2}(c) and \ref{fig minko 2}(d) show that
$\chi / V$ is negative in the density region of the intermediate phases,
even if we take into account the imperfection of the obtained nuclear shapes
and the uncertainties of the definition of the nuclear surface.
This means that the intermediate phases consist of nuclear surfaces
which are saddle-like at each point on average
and they consist of each highly connected nuclear and gas regions
due to a lot of tunnels [see Eq. (\ref{euler})].
Using the quantities $\langle H \rangle$ and $\chi/V$,
the sequence of the nuclear shapes with increasing the density
can be described as follows:
$(\langle H \rangle > 0,\ \chi/V > 0)$ $\rightarrow$
$(\langle H \rangle > 0,\ \chi/V = 0)$ $\rightarrow$
$(\langle H \rangle > 0,\ \chi/V < 0)$ $\rightarrow$
$(\langle H \rangle = 0,\ \chi/V = 0)$ $\rightarrow$
$(\langle H \rangle < 0,\ \chi/V < 0)$ $\rightarrow$
$(\langle H \rangle < 0,\ \chi/V = 0)$ $\rightarrow$
$(\langle H \rangle < 0,\ \chi/V > 0)$ $\rightarrow$ uniform.

Let us now consider the discrepancy from the results of previous works
which do not assume nuclear structure \cite{williams,lassaut};
the intermediate phases can not be seen in these works.
We can give following two reasons for the discrepancy.
\begin{itemize}
  \item[(1)] These previous calculations are based on the Thomas-Fermi approximation
    which can not sufficiently incorporate fluctuations of nucleon distributions.
    This shortcoming may result in favoring nuclei of smoothed simple shapes
    than in the real situation.
  \item[(2)] There is a strong possibility that some highly connected structures
    which have two or more substructures in a period
    are neglected in these works
    because only one structure is contained in a simulation box.
\end{itemize}

It is not unnatural that the phases with highly connected
nuclear and bubble regions are realized
as the most energetically stable state \cite{magierski,note rpw}.
It is considered that, for example,
a phase with perforated slablike nuclei, which has negative $\chi / V$,
could be more energetically stable than that with
extremely thin slablike nuclei.
The thin planar nucleus costs surface-surface energy
which stems from the fact that nucleons
bound in the nucleus feel its surfaces of both sides.
The surface-surface energy brings about an extra energy increase
in addition to the contribution of the surface energy.
We have to examine the existence of the intermediate phases
by more extensive simulations with larger nucleon numbers
and with longer relaxation time scales in the future.

%%%%%
\subsubsection{Minkowski functionals for $x=0.1$\label{minko n-rich}}

In the case of $x=0.1$, the criterion for identification of
the isodensity surface corresponding to the nuclear surface
using the second derivative of $V(\rho_{p, {\rm th}})/A(\rho_{p, {\rm th}})$
does not work at higher densities.
We thus use another method to calculate the normalized Minkowski functionals
of the nuclear surface for $x=0.1$.

\begin{figure}
\resizebox{7.0cm}{!}
{\includegraphics{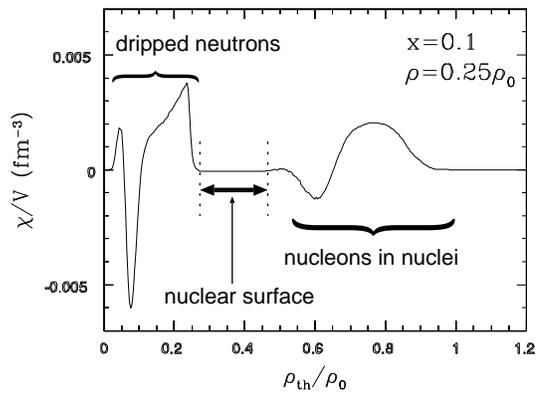}}
\caption{\label{fig euler typical}
  Euler characteristic density as a function of $\rho_{\rm th}$
  calculated for $x=0.1$ and $\rho=0.25\rho_{0}$.
  The contribution of the nuclear surface can be observed
  as the plateau region.
  }
\end{figure}

\begin{figure}
\resizebox{9.0cm}{!}
{\includegraphics{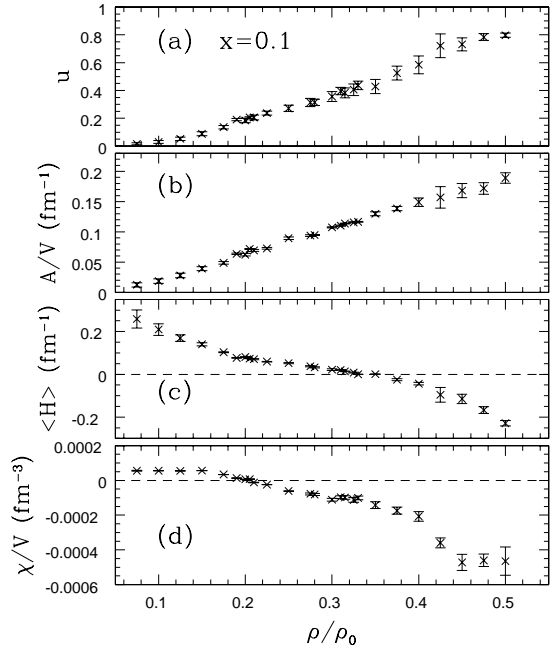}}
\caption{\label{fig minko x0.1}
  Density dependence of the normalized Minkowski functionals
  for cold matter at $x=0.1$.
  }
\end{figure}

In Fig.\ \ref{fig euler typical}, we have plotted the $\rho_{\rm th}$ dependence
of the Euler characteristic density $\chi/V$ at $\rho=0.25\rho_{0}$
as an example.  We can see that this curve consists of three components:
the peaks of the lower $\rho_{\rm th}$ region, the plateau region
and the peaks of the higher $\rho_{\rm th}$ region,
which are due to dripped neutrons (thus these peaks cannot be observed
for $x=0.5$), nuclear surfaces and nucleons in nuclei, respectively.
These components can also be seen at the other values of $\rho$
lower than $\rho_{\rm m}$.  However, we have to mention that
the higher the density becomes, the smaller the plateau region gets,
which means that the density contrast between the dripped neutron gas region
and the nuclear matter region becomes obscure.
Here, we take the mean values of the normalized Minkowski functionals
in the plateau region as those for the nuclear surface,
which are plotted as crosses in Fig.\ \ref{fig minko x0.1}.
The error bars shown in this figure are the standard deviations
of these quantities in the plateau region.
Consistency between this method and the one
using the second derivative of $V(\rho_{p, {\rm th}})/A(\rho_{p, {\rm th}})$
has been confirmed for $x=0.3$.

Fig.\ \ref{fig minko x0.1} shows
the resultant normalized Minkowski functionals for the nuclear surface
at various values of $\rho$ below $\rho_{\rm m}$.  The qualitative behaviors
of $u$ and $\langle H \rangle$ for $x=0.1$ are the same as those
for $x=0.5$ and 0.3; as $\rho$ increases,
$u$ increases and $\langle H \rangle$ decreases
(from positive to negative) almost monotonically in the density region of
0.1$\rho_{0} \alt \rho \alt \rho_{\rm m}$.
However, in the behaviors of $A/V$ and $\chi/V$,
qualitative differences can be observed between the present case
and the cases of $x=0.5$ and 0.3.
As can be seen in Fig.\ \ref{fig minko x0.1}(b),
$A/V$ increases almost linearly until just below $\rho_{\rm m}$.

The absence of the phases with cylindrical bubbles and with spherical bubbles
in the phase diagram of $x=0.1$ (Fig.\ \ref{phase diagram 2})
is well characterized by the behavior of $\chi/V$
shown in Fig.\ \ref{fig minko x0.1}(d).
In the cases of $x=0.5$ and 0.3, $\chi/V$ increases from negative to positive
with increasing density in the density region higher than that
of the slab phase. However, for $x=0.1$, we cannot observe the tendency
that $\chi/V$ starts increasing even at just below $\rho_{\rm m}$.
As a result, $\chi/V$ keeps negative until matter becomes uniform.
(Further discussion will be given in Section \ref{int}.)

Let us then consider the phase with slablike nuclei,
which have not been obtained in the simulations for $x=0.1$.
If it were realized by using a longer relaxation time scale,
it is expected to be obtained at $\rho \simeq 0.32$ -- 0.34 $\rho_{0}$
according to the behaviors of the $\langle H \rangle$ and $\chi/V$.
However, we cannot see any signs from Fig.\ \ref{fig minko x0.1}(b)
that $A/V$ stops increasing in this density region
unlike the behaviors of $A/V$ in the cases of $x=0.5$ and 0.3.
Here, we would like to mention that,
according to the Landau-Peierls argument,
thermal fluctuations are effective at destroying
the long-range order of one-dimensional layered lattice of slablike nuclei
rather than that of triangular lattice of rodlike nuclei and
of bcc lattice of spherical ones.
Thus, the melting temperature of the planar phase
would be lower than the other phases, which leads to a longer time scale
for formation of the slablike nuclei by the thermal diffusion.
Therefore, a further investigation with a longer relaxation time scale
is necessary to determine whether or not the phase with slablilke nuclei
is really prohibited in such neutron-rich matter in the present model.

\begin{figure}
\rotatebox{0}{
\resizebox{8.2cm}{!}
{\includegraphics{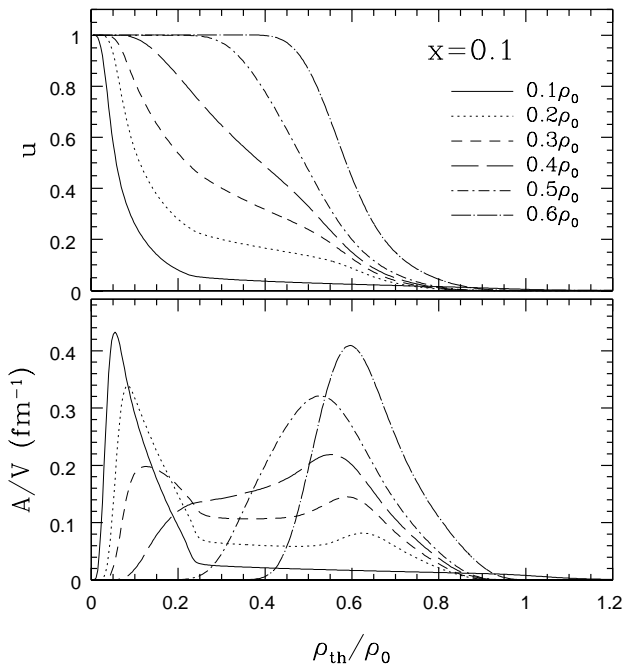}}}%
\caption{\label{fig minko thres x0.1}
  Volume fraction $u$ (upper panel) and surface area density $A/V$
  (lower panel) as functions of the threshold density $\rho_{\rm th}$
  calculated for $x=0.1$ and various nucleon densities $\rho$.
  }
%\end{center}
\end{figure}

In Fig.\ \ref{fig minko thres x0.1}, we have also plotted
$u$ and $A/V$ for the density distribution $\rho({\bf r})$
as functions of $\rho_{\rm th}$ like Figs.\ \ref{fig minko thres x0.5}
and \ref{fig minko thres x0.3}.
In comparison with Fig.\ \ref{fig minko thres x0.3},
the contribution of the dripped neutrons is shown more clearly
in this case.  We can see that the peak in the lower $\rho_{\rm th}$ region
due to the dripped neutrons combines to the peak in the higher $\rho_{\rm th}$
region.  This behavior stems from the fact that a part of the dripped neutrons
at lower densities are absorbed into nuclear matter region with increasing
the density at fixed $x$; finally, all the neutrons are contained there
in the uniform phase.
We can also expect from the $\rho_{\rm th}$ dependence of $A/V$
that the phase with slablike nuclei might be obtained around
$0.3 \rho_{0}$, where the slope of the plateau region is close to zero.

\section{Properties of the Effective Nuclear Interaction\label{int}}

Let us here examine the effective nuclear interaction used in this work.
Structure of matter at subnuclear densities is affected by the properties
of neutron-rich nuclei and of the pure neutron gas resulted from
the nuclear interaction.
Key quantities are the energy per nucleon $\epsilon_{n}$ of
the pure neutron matter, the proton chemical potential $\mu_{p}^{(0)}$
in the pure neutron matter, and the nuclear surface tension $E_{\rm surf}$.

There is a tendency, especially in the case of neutron star matter,
that the higher $\epsilon_{n}$,
the density $\rho_{\rm m}$ at which matter becomes uniform is lowered.
This is because larger $\epsilon_{n}$ tends to favor
uniform nuclear matter without dripped neutron gas regions
than mixed phases with dripped neutron gas regions.
In the neutron star matter, there is also a tendency that
the lower $\mu_{p}^{(0)}$, the smaller $\rho_{\rm m}$.
This is because $-\mu_{p}^{(0)}$ represents the
degree to which the neutron gas outside the nuclei
favors the presence of protons in itself.
The quantity $E_{\rm surf}$ controls the size of the nuclei and bubbles,
and hence the sum of the Coulomb and surface energies.
With increasing $E_{\rm surf}$ and so this energy sum,
$\rho_{\rm m}$ gets lowered.

It is important to check whether the effective nuclear force
given by Eqs.\ (\ref{hamiltonian})--(\ref{coulomb})
yields unrealistic values of these quantities or not.
If $\epsilon_{n}$, $|\mu_{p}^{(0)}|$ and $E_{\rm surf}$ for the present model
are unrealistically small in comparison with those for the other models,
our results which have reproduced the ``pasta'' phases
might be quite limited for the present model Hamiltonian.

\begin{figure}[htbp]
\rotatebox{0}{
\resizebox{8.2cm}{!}
{\includegraphics{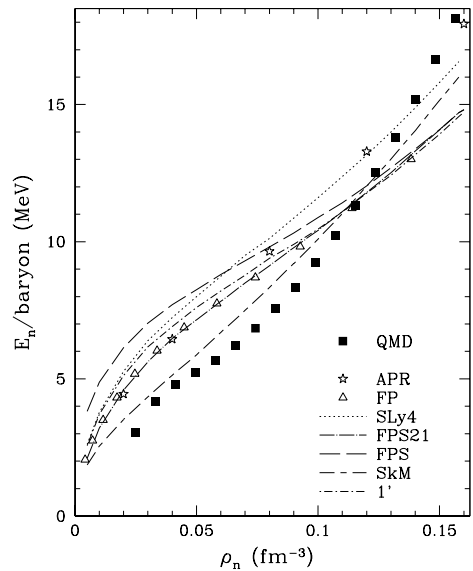}}}%
\caption{\label{fig e_n}
  The neutron density $\rho_{n}$ dependence of
  the energy per nucleon $\epsilon_{n}$ of the pure neutron matter.
  The solid squares show the result of the present QMD model Hamiltonian
  \cite{maruyama}.
  The dotted line denoted by SLy4 is the result from Ref.\ \cite{douchin}
  and the broken lines as marked by the other Skyrme interactions
  (FPS21, 1', FPS and SkM) are the results summarized by
  Pethick, Ravenhall and Lorenz \cite{pethick}.
  The open stars and triangles denote the values
  obtained by Akmal, Pandharipande and Ravenhall \cite{akmal},
  and by Friedman and Pandharipande \cite{fp}, respectively.
  }
\end{figure}

In order to evaluate $\epsilon_{n}$,
we set 1372 neutrons in a simulation box
imposed of the periodic boundary condition.
This system is cooled down by the QMD equations of motion with friction terms
[see Eqs.\ (\ref{qmdeom fric})] until the temperature becomes $\sim$ 1 keV.
The resultant values of $\epsilon_{n}$ are plotted in Fig.\ \ref{fig e_n}.
We note that our results for $\rho_{n}=$0.2, 0.6 and
1.0 $\rho_{0}$ (the result for $\rho_{n}=1.0 \rho_{0}$
is not plotted in Fig.\ \ref{fig e_n})
coincide with the results for zero-proton ratio
plotted in Fig.\ 9 of Ref.\ \cite{maruyama}.

The values of $\epsilon_{n}$ for the present model
behave like those for the SkM Skyrme interaction
especially in the density region of $\rho_{n} \alt 0.13$ fm$^{-3}$;
they are close to the result of the variational chain summation
obtained by Akmal, Pandharipande and Ravenhall \cite{akmal}
at $\rho_{n} \simeq \rho_{0}$.
The steep rise in $\epsilon_{n}$ in the higher neutron density region
($\rho_{n} \agt 0.1$ fm$^{-3}$)
compared to those obtained from the Hartree-Fock theory
using various Skyrme interactions
would help neutron-rich matter, which have larger dripped neutron density
of $\rho_{n} \agt 0.1$ fm$^{-3}$, to be uniform.
Therefore, we can say that this behavior of $\epsilon_{n}$
for the present QMD model Hamiltonian suppresses
the density region in which the ``pasta'' phases
are the most energetically favorable in neutron star matter
and in the case of $x=0.1$ in the present study.
We also note that $\epsilon_{n}$ at lower neutron densities of 
$\rho_{n} \alt 0.1$ fm$^{-3}$ is relatively small.
This, in turn, would lead to increase the density at which
matter with lower dripped neutron density
(e.g., the case of $x=0.3$ in the present study)
turns into uniform.

Next, we calculate the proton chemical potential $\mu_{p}^{(0)}$
in the pure neutron matter.
We use the cold neutron matter
prepared for the above calculation of $\epsilon_{n}$ as an initial condition.
We insert a proton into this pure neutron matter,
and then minimize the total energy by the frictional relaxation method
with fixing the positions and momenta of the other neutrons.
The position of the inserted proton is chosen randomly in the simulation box,
and its momentum is chosen randomly from $P \le 30$ MeV$/c$.
We evaluate $\mu_{p}^{(0)}$ as the difference in the total energy
between that before the insertion of the proton
and that after the optimization of the position and the momentum of the proton.

\begin{figure}[htbp]
\rotatebox{0}{
\resizebox{8.2cm}{!}
{\includegraphics{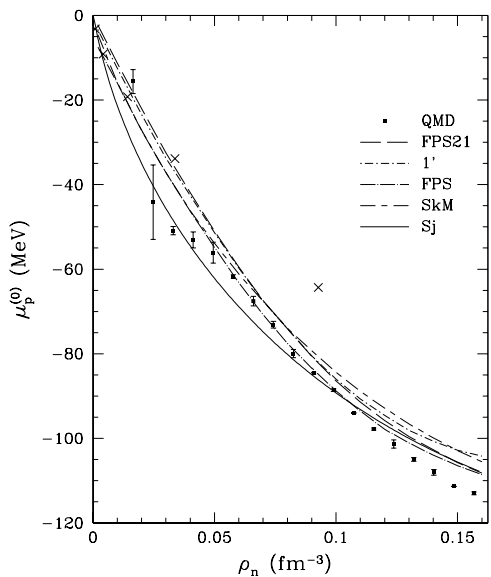}}}
\caption{\label{fig mu_p0}
  The neutron density dependence of
  the proton chemical potential $\mu_{p}^{(0)}$ in the pure neutron matter.
  The solid squares show the result of the present QMD Hamiltonian
  \cite{maruyama}.
  The broken lines as marked by the Skyrme interactions
  (FPS21, 1', FPS and SkM) are the results summarized by
  Pethick, Ravenhall and Lorenz \cite{pethick},
  and the solid line is the result of Sj\"oberg \cite{sj}.
  The crosses denote
  the values obtained by Siemens and Pandharipande
  \cite{siemens}.
  }
\end{figure}

In Fig.\ \ref{fig mu_p0}, we plot $\mu_{p}^{(0)}$
for the present model Hamiltonian.
As can be seen from this figure, the result for the present model Hamiltonian
generally reproduce the data of the other results
obtained from the Hartree-Fock theory using
the various Skyrme interactions
at densities $\rho \alt 0.1$ fm$^{-3}$.
At lower densities of $\rho \alt 0.025$ fm$^{-3}$,
errors are quite large and data scatter significantly.
This is because density fluctuations in pure neutron matter obtained by QMD
would be unrealistically large at such low densities
due to the fixed width of the wave packets in this model.
However, it is noted that even in such a density region,
our data are generally consistent with the other results mentioned above.

Finally, we turn to the surface tension,
which affects energetically favorable nuclear shape most directly
among the three quantities discussed here.
We have calculated $E_{\rm surf}$ from the energy change
induced by the change in the area of planar nuclear matter.
We use nucleon gas with proton fractions $x=0.1$, 0.2, 0.3 and 0.5
composed of 1372 nucleons.
In the calculation of $E_{\rm surf}$, the Coulomb interaction
is excluded.

In order to prepare slablike nuclear matter,
we first cool down the above nucleon gas from $k_{\rm B}T \sim 20$ MeV
to $\sim 0.2$ MeV using Eqs.\ (\ref{qmdeom fric})
in a shallow trapping harmonic potential
\begin{equation}
  V(z) = \frac{k}{2} z^{2}\ ,\label{trap}
\end{equation}
where $k=0.01$ MeV fm$^{-2}$.
The simulation box here is imposed of the periodic boundary condition
in the $x$- and $y$-directions, and is imposed of the open boundary condition
in the $z$-direction.
The box size $L_{x,y}$ in the $x$- and $y$-directions is set 20.26 fm.

\begin{figure}[htbp]
\rotatebox{0}{
\resizebox{8.2cm}{!}
{\includegraphics{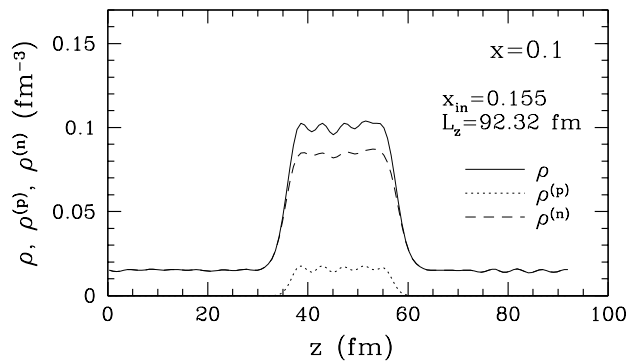}}}
\caption{\label{fig profile}
  The nucleon density profiles
  projected on the $z$-axis for the sample 1 of $x=0.1$.
  (The $z$-coordinate is shifted.)
  }
\end{figure}

When the temperature reaches $\sim 0.2$ MeV,
we remove the trapping potential and, except for the case with $x=0.5$,
we change the boundary condition in the $z$-direction from open to periodic.
The box size in the $z$-direction $L_{z}$ is chosen so as to
at least all nucleons dripped outside the nuclear matter region
can be contained in the box: $L_{z}= 92.32$ fm ($x=0.1$), 79.12 fm ($x=0.2$),
and 82.85 fm ($x=0.3$).
After we relax the system for $\sim$ 7000 fm/$c$,
we prepare three kinds of samples:
one is nothing changed (sample 1) and the others have the area of the $xy$ side
of the simulation box increased (decreased) by 1 $\%$
with the total volume of the box kept constant [sample 2 (sample 3)].
We further cool them down until $k_{\rm B}T \sim 0.1$ keV.
The resultant nucleon density profiles for the sample 1 of $x=0.1$
projected on the $z$-axis is shown in Fig.\ \ref{fig profile}.
As can be seen from this figure, $L_{z}=92.32$ fm is much larger than
the thickness of the slab $d \simeq 20$ fm,
and thus the volume of the dripped neutron gas region
is almost the same among the three kinds of samples
because the volume change in the nuclear matter region is negligible.
It is also noted that $d$ is much larger than the surface thickness
$d_{\rm surf}\simeq 5$ fm, which ensures that the surfaces at both sides
of the slab are separated well.
Therefore, we can say that the energy difference between
the sample 2 and 3 is just due to the difference in the surface area
of the planar nuclear matter.
Following the spirit of Ravenhall, Bennett and Pethick
(hereafter RBP) \cite{rbp}
and by using the sample 1,
we define the proton fraction $x_{\rm in}$ in the nuclear matter region
as an averaged value for the region of the width of $\simeq 5$ fm
in the central part of the slab,
where the proton and neutron density profiles ripple around constant values.

\begin{figure}[htbp]
\rotatebox{0}{
\resizebox{8.2cm}{!}
{\includegraphics{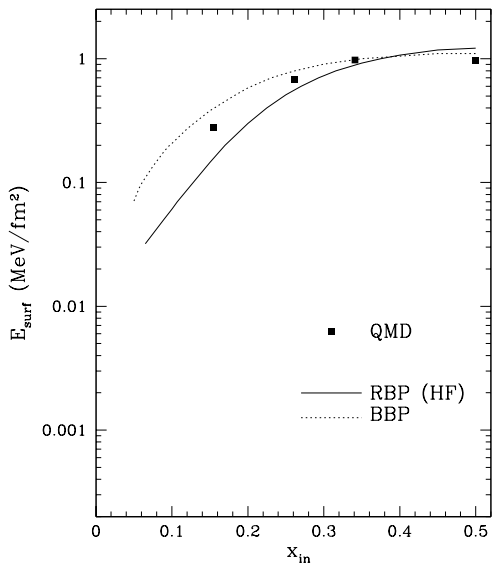}}}
\caption{\label{fig esurf}
  The nuclear surface energy
  per unit area (the surface tension) versus
  the proton fraction $x_{\rm in}$ in the nuclear matter region.
  The solid squares are the values of the present QMD Hamiltonian
  \cite{maruyama},
  the solid curve is the RBP result from their Hartree-Fock calculations
 \cite{rbp}, and the dotted curve is the BBP result \cite{bbp}.
  }
\end{figure}

We extract $E_{\rm surf}$ from the total energy $E_{2}$ of the sample 2
and $E_{3}$ of the sample 3 as follows:
\begin{equation}
  E_{\rm surf} =  \frac{E_{2} - E_{3}}{2 (S_{2} - S_{3})}\ ,
\end{equation}
where $S_{2}\ (S_{3})$ is the area of the $xy$ side of the sample 2 (sample 3)
given by $20.26^{2} \times 1.01$ fm$^{2}$ ($20.26^{2} \times 0.99$ fm$^{2}$).
The factor 2 in the denominator represents, of course,
the contribution of the two sides of the planar nuclear matter.
As shown in Fig.\ \ref{fig esurf},
all the results plotted in this figure almost coincide with each other
at $x_{\rm in}=0.5$, where uncertainty in the nuclear surface tension
is rather small.
These results deviate significantly at lower $x_{\rm in}$,
and the values of $E_{\rm surf}$ of the present calculation lie between
those obtained by Baym, Bethe and Pethick (hereafter BBP) \cite{bbp}
and by RBP \cite{rbp}, which is based on the Hartree-Fock calculation
using a Skyrme interaction,
at $x_{\rm in} \sim 0.15 - 0.35$.
Thus we can say that, in comparison with the result by RBP
taken to be standard here,
the contribution of $E_{\rm surf}$
of the present QMD model tends to favor uniform nuclear matter
rather than the inhomogeneous ``pasta'' phases
for lower $x_{\rm in}$ of $\alt 3.5$.

Now let us discuss the fact that the intermediate phases
with a negative value of $\chi/V$ are obtained instead of the bubble phases 
in a wide density region for $x=0.1$.
In considering this problem, we should note the general tendency that
$|\mu_{p}^{(0)}|$ increases and $E_{\rm surf}$ decreases
as matter becomes neutron-rich
(see Figs.\ \ref{fig mu_p0} and \ref{fig esurf}).
The density dependence of $A/V$ monotonically increases until $\rho_{\rm m}$
as shown in Fig.\ \ref{fig minko x0.1} would partly stem from the small value
of $E_{\rm surf}$.  The apparently lower melting density
$\rho_{\rm m}$ in this case than in the cases of $x=0.5$ and 0.3,
even though $E_{\rm surf}$ is small, is due to a large $|\mu_{p}^{(0)}|$,
which increases typically of order 10 MeV as $\rho_{n}$ increases.
The small $E_{\rm surf}$ and the large $|\mu_{p}^{(0)}|$
in neutron-rich matter would help nuclear matter regions
and neutron gas regions mix each other at the cost of
small surface energy below $\rho_{\rm m}$.
As a result, the structures with a negative $\chi/V$ could be favored.
According to Fig.\ \ref{fig mu_p0}, the quantity $\mu_{p}^{(0)}$ obtained
for the present model Hamiltonian is consistent with those for the other
Skyrme-Hartree-Fock calculations in the relevant region of
$\rho_{n} \alt 0.08$ fm$^{-3}$.  It is thus possible that
a result which shows the structure of matter changes from
negative $\chi/V$ to uniform without undergoing ``swiss cheese'' structure
will be obtained by another calculation for neutron-rich matter
using some framework without assuming nuclear shape.

In closing this subsection, we summarize the consequences of the resultant
$\epsilon_{n}$, $\mu_{p}^{(0)}$ and $E_{\rm surf}$
for the present QMD Hamiltonian.
\begin{itemize}
  \item[1.] For symmetric matter ($x=x_{\rm in}=0.5$)
    
    According to $E_{\rm surf}$ at $x_{\rm in}=0.5$,
    the present model is consistent with the other results,
    and is an appropriate effective interaction
    for the study of the ``pasta'' phases at $x=0.5$.

  \item[2.] For neutron-rich cases

    For neutron-rich cases such as $x \sim 0.1$,
    in which the dripped neutron density $\rho_{n}$ grows
    $\agt 0.1$ fm$^{-3}$ just before matter turns into uniform,
    the present QMD model can be taken as a conservative one
    in reproducing the ``pasta'' phases.
    Its $\epsilon_{n}$, $\mu_{p}^{(0)}$ and $E_{\rm surf}$ act to suppress
    the density region of the ``pasta'' phases compared to
    other Skyrme-Hartree-Fock results.

  \item[3.] For intermediate cases

    At intermediate proton fraction of $x \sim 0.3$,
    $\epsilon_{n}$ of the present model acts to favor
    the inhomogeneous ``pasta''phases
    rather than the uniform phase and $E_{\rm surf}$
    acts in the opposite way
    in comparison with other Skyrme-Hartree-Fock results.
\end{itemize}

\section{Astrophysical Discussions}

Here we would like to discuss astrophysical consequences of our results.
Pethick and Potekhin have pointed out that
elastic properties of ``pasta'' phases with rodlike and slablike nuclei
are similar to those of liquid crystals, which stems from
the similarity in the geometrical structures \cite{pp}.
It can also be said that the intermediate phases
observed in the present work are ``spongelike''
(or ``rubberlike'' for $\langle H \rangle < 0$) phases
because they have both highly connected nuclear and bubble regions
shown as $\chi / V < 0$.
The elastic properties of the spongelike intermediate phases
are qualitatively different from
those of the liquid-crystal-like ``pasta'' phases
because the former ones do not have any directions
in which the restoring force does not act;
while the latter ones have.
Our results imply that the intermediate phases
occupy a significant fraction of the density region
in which nonspherical nuclei can be seen (see Figs.\ \ref{phase diagram 1}
and \ref{phase diagram 2}).
According to Figs.\  \ref{phase diagram 1} and \ref{phase diagram 2},
we expect that the maximum elastic energy that can be stored
in the neutron star crust and supernova inner core is higher than
that in the case where all nonspherical nuclei have
simple ``pasta'' structures.
Besides, the cylinder and the slab phases,
which are liquid-crystal-like, lie between
the spongelike intermediate phases or the crystalline solidlike phase,
and the releasing of the strain energy would, in consequence,
concentrate in the domain of these liquid-crystal-like phases.
The above effects of the intermediate phases
should be taken into account
in considering the crust dynamics of starquakes
and hydrodynamics of the core collapse, etc.
if these phases exist in neutron star matter and supernova matter.
In the context of pulsar glitch phenomena,
the effects of the spongelike nuclei
on the pinning rate and the creep velocity of superfluid neutron vortices
also have yet to be investigated.

For neutrino cooling of neutron stars,
some version of the direct URCA process
which is suggested by Lorenz et al. \cite{lorenz},
that this might be allowed in the ``pasta'' phases,
would be suppressed in the intermediate phases.
This is due to the fact that the proton spectrum at the Fermi surface
is no longer continuous in the spongelike nuclei.
An important topic which we would like to mention is
about the effects of the intermediate phases on neutrino trapping
in supernova cores.
The nuclear parts connect over a wide region
which is much larger than that
characterized by the typical neutrino wave length $\sim 20$ fm.
Thus the neutrino scattering processes are no longer coherent
in contrast to the case of the spherical nuclei,
and this may, in consequence, reduce the diffusion time scale of neutrinos
as in the case of ``pasta'' phases with simple structures.
This reduction softens the supernova matter
and would thus act to enhance the amount of the released gravitational energy.
It would be interesting to estimate the neutrino opacity
of the spongelike phases and the ``pasta'' phases.

Finally, we would like to mention the
thermal fluctuations with long wavelengths
leading to displacements of ``pasta'' nuclei,
which cannot be incorporated into the simulations
using a finite-size box.
Even if we succeed in reproducing the phase with slablike nuclei
in neutron-rich matter in the future study,
we should remind the above effect of thermal fluctuations
to consider the real situation of matter in inner crusts of neutron stars.
Following the discussions in Refs.\ \cite{gentaro1,gentaro2,gentaro thesis}
by a liquid-drop model, it is likely that the extension of slablike nuclei
is limited to a finite length scale of $\sim O(10^{2}-10^{3})$ fm
in the temperature regions typical for neutron star crusts
and supernova cores.

\section{Summary and Conclusion}

We have performed QMD simulations for matter
with fixed proton fractions $x=0.5$, 0.3 and 0.1 at various densities
below the normal nuclear density.
Our calculations without any assumptions on the nuclear shape
demonstrate that the ``pasta'' phases with rodlike nuclei,
with slablike nuclei, with cylindrical bubbles and with spherical bubbles
can be formed dynamically from hot uniform matter
within the time scale of $\tau \sim O(10^{3}-10^{4})$ fm/$c$
in the proton-rich cases of $x=0.5$ and 0.3.
We also demonstrate that the ``pasta'' phase with cylindrical nuclei
can be formed dynamically within the time scale of $\tau \sim O(10^{4})$ fm/$c$
for the neutron-rich case of $x=0.1$.
Our results imply the existence of at least the phase with cylindrical nuclei
in neutron star crusts
because they cool down keeping the local thermal equilibrium
after proto-neutron stars are formed
and their cooling time scale, which is macroscopic one,
is much larger than the relaxation time scale of our simulations.

In addition to these ``pasta'' phases with simple structures,
our results obtained here also suggest the existence of
intermediate phases which have complicated nuclear shapes.
We have systematically analyzed the structure of matter
with two-point correlation functions
and with morphological measures ``Minkowski functionals'',
and have demonstrated how structure changes with increasing density.
Making use of a topological quantity called Euler characteristic,
which is one of the Minkowski functionals,
it has been found that the intermediate phases can be characterized
as those with negative Euler characteristic.
This means that the intermediate phases have ``spongelike''
(or ``rubberlike'' for $\langle H \rangle < 0$) structures
which have both highly connected nuclear and bubble regions.
The elastic properties of the spongelike intermediate phases
are qualitatively different from
those of the liquid-crystal-like ``pasta'' phases.

We have also investigated the properties of the effective QMD interaction
used in the present work in order to examine the validity of our results.
Important quantities which affect the structure of matter
are the energy per nucleon $\epsilon_{n}$ of the pure neutron matter,
the proton chemical potential $\mu_{p}^{(0)}$ in pure neutron matter and
the nuclear surface tension $E_{\rm surf}$.
The resultant these quantities
show that the present QMD interaction has generally reasonable properties
at subnuclear densities among other nuclear interactions.
It is thus concluded that our results are not exceptional ones
in terms of nuclear forces.

Our results which suggest the existence of the highly connected
intermediate phases as well as the simple ``pasta'' phases
provide a vivid picture that matter in neutron star inner crusts
and supernova inner cores has a variety of material phases.
The stellar region which we have tried to understand throughout this paper
is relatively tiny, but there has quite rich properties which stem from
the fancy structures of dense matter.

\begin{acknowledgments}
G. W. is grateful to T. Maruyama, K. Iida,
A. Tohsaki, H. Horiuchi, K. Oyamatsu, H. Takemoto,
K. Niita, S. Chikazumi, I. Kayo,
C. Hikage, T. Buchert, N. Itoh and K. Kotake
for helpful discussions and comments.
G. W. also appreciate J. Lee for checking the manuscript.
This work was supported in part
by the Junior Research Associate Program in RIKEN
through Research Grant No. J130026 and
by Grants-in-Aid for Scientific Research
provided by the Ministry of
Education, Culture, Sports, Science and Technology
through Research Grant (S) No. 14102004, No. 14079202 and No. 14-7939.
\end{acknowledgments}

\appendix
\section{The Ewald Sum for Particles with a Gaussian Charge Distribution
  \label{ewald sum}}

In calculating a long-range interaction, such as the Coulomb interaction,
it is necessary to sum up all contributions of particles
at a sufficiently far distance.
The Ewald sum is a familiar technique for efficiently computing
the long-range contributions in a system with the periodic boundary condition
(see, e.g., Refs.\ \cite{allen,tosi}; recent mathematically careful discussion
relating to the conditionally convergence of the Coulomb energy
can be seen, e.g., in Ref.\ \cite{takemoto}).
The basic idea of the Ewald sum is that
the contributions of particles in a long distance in real space
can be calculated as contributions in the neighborhood
in Fourier space:
the contributions of particles in a short distance is summed up in real space
and those of particles in a long distance is summed up in Fourier space.

\begin{figure}[htbp]
\begin{center}
\rotatebox{270}{
\resizebox{3.5cm}{!}
{\includegraphics{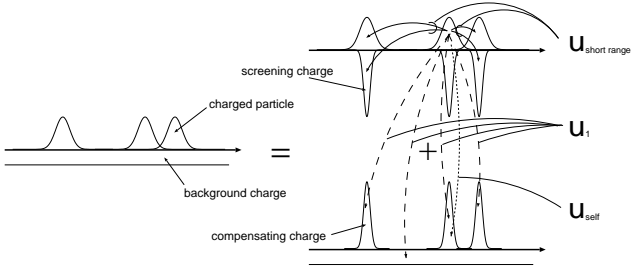}}}
\caption{\label{fig ewald}
  Schematic picture of charge distribution in the Ewald sum
  for particles with a Gaussian charge distribution.
  }
\end{center}
\end{figure}

Let us consider a system consisting of charged particles
which have a Gaussian charge distribution
and a uniform background charge
which cancels the total charge of charged particles.
These $N$ particles are assumed to be in a cubic simulation box
with volume $V=L_{\rm box}^{3}$
which is imposed by the periodic boundary condition.

If every particle $i$ with total charge $Z_{i}$
is surrounded by a Gaussian charge distribution
with total charge $-Z_{i}$,
the electrostatic interaction of particle $i$
turns into a screened short-range interaction.
Thus the total Coulomb energy ${\cal U_{\rm Coul}}$ of this system
can be decomposed into as follows (see Fig.\ \ref{fig ewald}):
\begin{equation}
  {\cal U_{\rm Coul}} = {\cal U_{\rm short\ range}} + {\cal U}_{1}
  - {\cal U_{\rm self}}\ , \label{coul}
\end{equation}
where ${\cal U_{\rm short\ range}}$ is the sum of the Coulomb energy
between an each unscreened charged particle $i$
and the other charged particles with screening charges,
${\cal U}_{1}$ is that between an each unscreened charged particle $i$
and compensating charges which cancels the screening Gaussian charges,
and ${\cal U_{\rm self}}$ is the sum of spurious self interactions
between an each charged particle $i$ and its compensating charge $i$.

Charge densities of the real charged particles $\rho({\bf r})$
and of the screening charges $\rho_{s}({\bf r})$ can be written as
\begin{eqnarray}
  \rho({\bf r}) &=& \sum_{\bf n} \sum_{i=1}^{N}
  Z_{i} \left( \frac{a}{\pi} \right)^{3/2}
  e^{-a | {\bf r}-({\bf r}_{i} + L_{\rm box} {\bf n}) |^{2}}\nonumber\\
  &\equiv& \sum_{\bf n} \sum_{i=1}^{N} \rho_{({\bf n}, i)}({\bf r}), \\
  \rho_{s}({\bf r}) &=& -\sum_{\bf n} \sum_{i=1}^{N}
  Z_{i} \left( \frac{\alpha_{\rm Ewald}}{\pi} \right)^{3/2}
  e^{-\alpha_{\rm Ewald} | {\bf r}-({\bf r}_{i} + L_{\rm box} {\bf n}) |^{2}}
  \nonumber\\
  &\equiv& \sum_{\bf n} \sum_{i=1}^{N} \rho_{s, ({\bf n}, i)}({\bf r}),
\end{eqnarray}
where $a$ and $\alpha_{\rm Ewald}$ are the reciprocals of the widths
of charge distributions for a charged particle and a screening charge,
respectively,
and {\bf n} denotes a position vector of a periodic image
normalized by $L_{\rm box}$.
Thus a distribution of screened charges $\rho_{\rm screened}$ is
\begin{equation}
  \rho_{\rm screened}({\bf r}) \equiv \rho({\bf r}) + \rho_{s}({\bf r})\ .
\end{equation}
The electrostatic potential $\phi_{\rm short\ range}({\bf r})$
due to $\rho_{\rm screened}({\bf r})$ can be obtained as
\begin{widetext}
\begin{equation}
  \phi_{\rm short\ range}({\bf r}) = \sum_{\bf n} \sum_{i=1}^{N} Z_{i}
  \left\{ -\frac{{\rm erfc}(\sqrt{a}\ | {\bf r}-({\bf r}_{i}+L_{\rm box} {\bf n})|)}
      {|{\bf r}-({\bf r}_{i}+L_{\rm box} {\bf n})|}
      + \frac{{\rm erfc}(\sqrt{\alpha_{\rm Ewald}}\
        | {\bf r}-({\bf r}_{i}+L_{\rm box} {\bf n})|)}
      {|{\bf r}-({\bf r}_{i}+L_{\rm box} {\bf n})|}
      \right\} +\ C\ ,
\end{equation}
\end{widetext}
because a solution of the Poisson equation
\begin{equation}
  -\frac{1}{r} \frac{d^{2}}{dr^{2}}(r \phi(r))
  = 4 \pi \left[ Z \left( \frac{a}{\pi} \right)^{3/2} e^{-ar^{2}} \right]
\end{equation}
is
\begin{equation}
  \phi(r) = Z \frac{{\rm erf}(\sqrt{a}\ r)}{r} + {\rm const.}\ ,\label{solution}
\end{equation}
where erf($x$) and erfc($x$) are the error function
and the complementary error function, respectively,
and they are defined as
$\ {\rm erf}(x) \equiv \frac{2}{\sqrt{\pi}} \int_{0}^{x} \exp{(-s^{2})}\ ds\ $
and $\ {\rm erfc}(x) \equiv 1- {\rm erf}(x)$.
Here, we determine the constant $C$
so that the average value of $\phi_{\rm short\ range}$ in the simulation box $V$
be zero:
\begin{eqnarray}
  &&\int_{V} \phi_{\rm short\ range}({\bf r})\ d^{3}{\bf r}\nonumber\\
  &=& \sum_{i=1}^{N} Z_{i}
  \left( -\frac{\pi}{a} + \frac{\pi}{\alpha_{\rm Ewald}} \right) + C V = 0\ .
\end{eqnarray}
Thus, $\phi_{\rm short\ range}({\bf r})$ leads to
\begin{widetext}
\begin{eqnarray}
  \phi_{\rm short\ range}({\bf r}) &=& \sum_{\bf n} \sum_{i=1}^{N} Z_{i}
  \left\{ -\frac{{\rm erfc}(\sqrt{a}\ | {\bf r}-({\bf r}_{i}+L_{\rm box} {\bf n})|)}
      {|{\bf r}-({\bf r}_{i}+L_{\rm box} {\bf n})|}
      + \frac{{\rm erfc}(\sqrt{\alpha_{\rm Ewald}}\
        | {\bf r}-({\bf r}_{i}+L_{\rm box} {\bf n})|)}
      {|{\bf r}-({\bf r}_{i}+L_{\rm box} {\bf n})|}
      \right\} \nonumber\\
      && - \left( -\frac{\pi}{a} + \frac{\pi}{\alpha_{\rm Ewald}} \right)
      \rho_{\rm avr}\ ,
\end{eqnarray}
\end{widetext}
where the average charge density $\rho_{\rm avr}$ of the charged particles
is defined as
\begin{equation}
  \rho_{\rm avr}\equiv\sum_{i=1}^{N} \frac{Z_{i}}{V}\ .
\end{equation}

The total Coulomb energy ${\cal U_{\rm short\ range}}$
between an each unscreened real charged particle $i$ and
the other charged particles with a screening charge can be calculated as
\begin{widetext}
\begin{eqnarray}
  {\cal U_{\rm short\ range}} &=&
  \frac{1}{2} \sum_{i=1}^{N} \Biggl\{
    \int d^{3}{\bf r} \int d^{3}{\bf r}'
    \rho_{({\bf n}=0, i)}({\bf r}) \nonumber\\&&\left.\quad
    \sum_{{\bf n}'} {\sum_{j}}' Z_{j}
    \left\{ -\frac{{\rm erfc}(\sqrt{a}\ | {\bf r}-({\bf r}_{j}+L_{\rm box} {\bf n})|)}
      {|{\bf r}-({\bf r}_{j}+L_{\rm box} {\bf n})|}
      + \frac{{\rm erfc}(\sqrt{\alpha_{\rm Ewald}}\
        | {\bf r}-({\bf r}_{j}+L_{\rm box} {\bf n})|)}
      {|{\bf r}-({\bf r}_{j}+L_{\rm box} {\bf n})|}
    \right\} \right.\nonumber\\&&\quad\quad
    - \left( -\frac{\pi}{a} + \frac{\pi}{\alpha_{\rm Ewald}} \right)
      \rho_{\rm avr} \Biggr\}\nonumber\\
      &=& \frac{1}{2} \sum_{{\bf n}} {\sum_{i,j}}'
    \frac{Z_{i} Z_{j}}{|{\bf r}_{i} - ({\bf r}_{j} + L_{\rm box} {\bf n})|}
    \left\{ {\rm erfc}\left( \sqrt{\frac{a\ \alpha_{\rm Ewald}}{a+\alpha_{\rm Ewald}}}\
          |{\bf r}_{i} - ({\bf r}_{j} + L_{\rm box}{\bf n})| \right) \right.
      \nonumber \\
        &&  \quad - {\rm erfc}\left( \sqrt{\frac{a}{2}}\
          |{\bf r}_{i} - ({\bf r}_{j} + L_{\rm box}{\bf n})| \right) \Biggr\}
        - \frac{V}{2}
        \left( -\frac{\pi}{a} + \frac{\pi}{\alpha_{\rm Ewald}} \right)
        \rho_{\rm avr}^{2}\ ,
      \label{u_short}
\end{eqnarray}
\end{widetext}
where the primes on the summations mean the terms $i=j$ at ${\bf n}=0$
are excluded.

Next, we calculate ${\cal U}_{1}$, which is the sum of the Coulomb energy
between an unscreened charged particle $i$
and a charge density $\rho_{1}({\bf r})$
which consists of the compensating charges and the background charge:
\begin{eqnarray}
  \rho_{1}({\bf r}) &=&
  -\ \rho_{s}({\bf r}) - \rho_{\rm avr} \nonumber \\
  &=& \sum_{\bf n} \sum_{j=1}^{N} Z_{j}
  \left( \frac{\alpha_{\rm Ewald}}{\pi} \right)^{3/2}
  e^{-\alpha_{\rm Ewald} |{\bf r}-({\bf r}_{j}+L_{\rm box} {\bf n})|^{2}}
  \nonumber\\
  && -\ \rho_{\rm avr}\ .
\end{eqnarray}
Fourier transforming the charge distribution $\rho_{1}({\bf r})$ yields
\begin{eqnarray}
  \rho_{1}({\bf k}) &=&
  \frac{1}{V} \int_{V} d^{3}{\bf r} e^{-i{\bf k}\cdot{\bf r}} \rho_{1}({\bf r})
  \nonumber \\
  &=& \frac{1}{V} \sum_{j=1}^{N} Z_{j} e^{-i{\bf k}\cdot{\bf r}_{j}}
  e^{-\frac{k^{2}}{4 \alpha_{\rm Ewald}}}
  - \rho_{\rm avr}\ \delta_{\bf k}\ .
\end{eqnarray}

Using $\rho_{1}({\bf k})$ and the Poisson equation
$(\nabla^{2} \phi_{1}({\bf r}) = - 4\pi \rho_{1}({\bf r}))$
in the Fourier form
\begin{equation}
  k^{2} \phi_{1}({\bf k}) = 4\pi \rho_{1}({\bf k})\ ,
\end{equation}
we can at once obtain the electrostatic potential $\phi_{1}$
due to the charge density $\rho_{1}$:
\begin{eqnarray}
  \phi_{1}({\bf r}) &=&
  \sum_{\bf k} \phi_{1}({\bf k}) e^{i{\bf k}\cdot{\bf r}} \nonumber \\
  &=& \frac{1}{V} \sum_{{\bf k} \neq 0} \sum_{j=1}^{N}
  \frac{4\pi Z_{j}}{k^{2}}\ e^{i{\bf k}\cdot ({\bf r}-{\bf r}_{j})}
  e^{-\frac{k^{2}}{4 \alpha_{\rm Ewald}}}\ .\quad
\end{eqnarray}
The term with ${\bf k}=0$ is canceled due to charge neutrality.
Thus the Coulomb energy ${\cal U}_{1}$ is given by
\begin{eqnarray}
  {\cal U}_{1} &=&
  \frac{1}{2} \sum_{i=1}^{N} \int d^{3}{\bf r}
  Z_{i} \left( \frac{a}{\pi} \right)^{3/2} e^{-a | {\bf r}-{\bf r}_{i}|^{2}}
  \phi_{1}({\bf r}) \nonumber \\
  &=& \frac{1}{2} \sum_{{\bf k} \neq 0} \sum_{i,j}
  \frac{4\pi Z_{i} Z_{j}}{V k^{2}}\
  e^{i{\bf k}\cdot ({\bf r}_{i}-{\bf r}_{j})} \nonumber\\
  &&\times\
  e^{-\frac{k^{2}}{4} \left( \frac{1}{\alpha_{\rm Ewald}}+\frac{1}{a} \right)}\ .
  \label{u1}
\end{eqnarray}

We have to subtract a sum of spurious self interactions
${\cal U}_{{\rm self,}i}$ between a charged particle $i$
and its compensating charge from ${\cal U}_{1}$.
According to Eq. (\ref{solution}),
an electrostatic potential $\phi_{\rm Gauss}$
due to a Gaussian compensating charge is
$\phi_{\rm Gauss}({\rm r}) = Z_{i}\ {\rm erf}(\sqrt{\alpha_{\rm Ewald}}\ r)/r$,
thus the self interaction of particle $i$ reads to
\begin{eqnarray}
  {\cal U}_{{\rm self,} i} &\equiv&
  \int Z_{i} \left( \frac{a}{\pi} \right)^{3/2}
  e^{-a r^{2}} \phi_{\rm Gauss}(r)\ d^{3}{\bf r}\nonumber\\
  &=& 2 \left( \frac{a}{\pi} \right)^{1/2} Z_{i}^{2}
  \sqrt{\frac{\alpha_{\rm Ewald}}{a + \alpha_{\rm Ewald}}}\ ,
\end{eqnarray}
and hence,
\begin{equation}
  {\cal U_{\rm self}} =
  \frac{1}{2} \sum_{i=1}^{N} {\cal U}_{{\rm self,} i}
  = \frac{1}{\sqrt{\pi}} \sqrt{\frac{a\ \alpha_{\rm Ewald}}{a + \alpha_{\rm Ewald}}}
    \sum_{i=1}^{N} Z_{i}^{2}\ . \label{u_self}
\end{equation}
Finally, the total Coulomb energy can be calculated by
Eqs.\ (\ref{coul}), (\ref{u_short}), (\ref{u1}) and (\ref{u_self}).
The positive background charge does not appear explicitly
because the average value of $\phi_{\rm short\ range}$
within the simulation box is set to be zero.

\begin{figure}[htbp]
\rotatebox{0}{
\resizebox{9.0cm}{!}
{\includegraphics{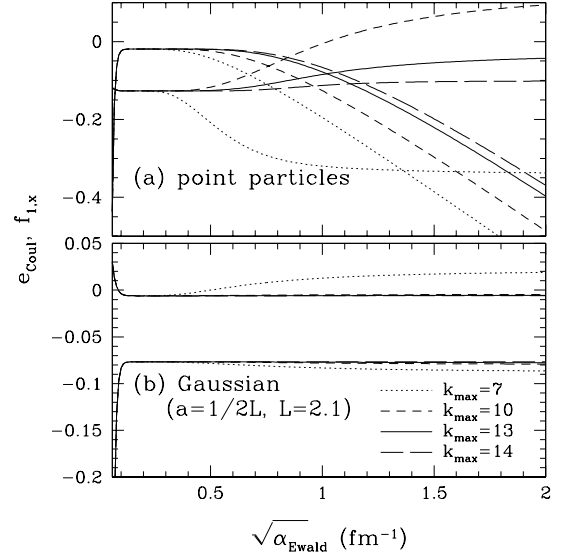}}}
\caption{\label{fig ewald}
  The total Coulomb energy per particle $e_{\rm Coul}$ (in units of MeV) and
  the $x$ component of the Coulomb force $f_{1,x}$ (in units of MeV/fm)
  acting on a particle
  obtained from Eqs.\ (\ref{coul}), (\ref{u_short}), (\ref{u1})
  and (\ref{u_self}) as a function of $\sqrt{\alpha_{\rm Ewald}}$.
  We use 1024 protons distributed randomly
  in a box of $L_{\rm box}=39.59$ fm (i.e. $\rho_{p}=0.1\rho_{0}$).
  The results shown in (a) and (b) are calculated for the same configuration
  of particle positions $\{ {\bf r}_{i}\}$ but for different values
  of the width $a$ of the Gaussian distribution; (a) $a=0$ (point charge) and
  (b) $a=1/2L$ with $L=2.1$ fm$^{2}$.
  }
\end{figure}

The total Coulomb energy per particle $e_{\rm Coul}$
and the $x$ component of the Coulomb force $f_{1,x}$ acting on a particle
for various values of $\alpha_{\rm Ewald}$
are plotted in Fig.\ \ref{fig ewald}. In this calculation,
we use 1024 positive charged particles (protons) distributed randomly
in a simulation box of $L_{\rm box}=39.59$ fm (i.e. $\rho_{p}=0.1\rho_{0}$),
which is imposed of the periodic boundary condition.
Figs.\ \ref{fig ewald} (a) and \ref{fig ewald} (b) show the results
for point charges and for Gaussian charge distributions, respectively,
which are calculated for the same configuration
of the particle positions $\{ {\bf r}_{i}\}$.
The width of the Gaussian charge distributions is set to be
$a=1/2L$ with $L=2.1$ fm$^{2}$, which corresponds to the width of the
wave packets in the QMD model used in this work.

We note that there are plateau regions of $e_{\rm Coul}$ and $f_{1,x}$
whose values do not depend on $\alpha_{\rm Ewald}$.
These constant values give results to be obtained.
We note that the range of $\sqrt{\alpha_{\rm Ewald}}$
of the plateau regions become larger
with increasing $k_{\rm max}$, where $k_{\rm max}$ is the cutoff radius
in the unit of $2\pi/L_{\rm box}$ for the summation in Fourier space.
As can be seen from Fig.\ \ref{fig ewald}, $\alpha_{\rm Ewald}$ dependences
of $e_{\rm Coul}$ and $f_{1,x}$ for the present QMD model
with a finite width of the Gaussian charge distributions are weaker
than those for the point charges.
These features are also confirmed for different proton number densities
of 0.2 and 0.3 $\rho_{0}$.
In our simulations, $\alpha_{\rm Ewald}$ is set 13 or 14,
which are considered to be large enough to calculate the total Coulomb energy
per particle in accuracy less than $O(1)$ keV,
which is the typical value of the energy difference
between successive ``pasta'' phases in neutron star matter
obtained by previous works.

%\bibliography{qmd1}

\begin{thebibliography}{99}
\bibitem{note nose} In some cases, the Nos\'e-Hoover-like thermostat
  which will be described in another paper \cite{qmd hot} is also used.
  In the present study, employment of this thermostat is almost limited
  in a higher temperature region where matter is completely uniform.
\end{thebibliography}

\end{document}